\documentclass[manuscript]{aastex61}
\usepackage{natbib}
\usepackage{amsmath}
\usepackage{graphicx}

\definecolor{orange}{rgb}{.8,0.4,0}

\shorttitle{Transmission Spectrum of GJ 1132b}
\shortauthors{Libby-Roberts et al.}

\begin{document}

\title{The Featureless HST/WFC3 Transmission Spectrum of the Rocky Exoplanet GJ 1132b: No Evidence For A Cloud-Free Primordial Atmosphere and Constraints on Starspot Contamination}

\correspondingauthor{Jessica E. Libby-Roberts}
\email{jessica.e.roberts@colorado.edu}

\author[0000-0002-2990-7613]{Jessica E. Libby-Roberts}
\affil{Department of Astrophysical and Planetary Sciences, University of Colorado, Boulder, CO 80309, USA}

\author[0000-0002-3321-4924]{Zachory K. Berta-Thompson}
\affil{Department of Astrophysical and Planetary Sciences, University of Colorado, Boulder, CO 80309, USA}

\author[0000-0001-8274-6639]{Hannah Diamond-Lowe}
\affil{National Space Institute, Technical University of Denmark, Elektrovej, 2800 Kgs.\ Lyngby, Denmark}

\author[0000-0002-4020-3457]{Michael A. Gully-Santiago}
\affil{The University of Texas at Austin Department of Astronomy, 2515 Speedway, Stop C1400, Austin, TX 78712, USA}

\author{Jonathan M. Irwin}
\affil{Center for Astrophysics, Harvard and Smithsonian, 60 Garden Street, Cambridge, MA, 02138, USA}

\author[0000-0002-1337-9051]{Eliza M.-R. Kempton}
\affil{Department of Astronomy, University of Maryland, College Park, MD 20742, USA}

\author[0000-0002-3627-1676]{Benjamin V.\ Rackham}
\altaffiliation{51 Pegasi b Fellow}
\affil{Department of Earth, Atmospheric and Planetary Sciences, and Kavli Institute for Astrophysics and Space Research, Massachusetts
Institute of Technology, Cambridge, MA 02139, USA}

\author[0000-0002-9003-484X]{David Charbonneau}
\affil{Center for Astrophysics, Harvard and Smithsonian, 60 Garden Street, Cambridge, MA, 02138, USA}

\author[0000-0002-0875-8401]{Jean-Michel D\'esert}
\affil{Anton Pannekoek Institute for Astronomy, University of Amsterdam, 1090 GE Amsterdam, The Netherlands}

\author[0000-0001-7730-2240]{Jason A. Dittmann}
\affil{Max Planck Institute for Astronomy, Königstuhl 17, 69117, Heidelberg, Germany}

\author{Ryan Hofmann}
\affil{Department of Astrophysical and Planetary Sciences, University of Colorado, Boulder, CO 80309, USA}
\affil{National Solar Observatory, Boulder, CO 80309, USA}

\author[0000-0002-4404-0456]{Caroline V. Morley}
\affil{The University of Texas at Austin Department of Astronomy, 2515 Speedway, Stop C1400, Austin, TX 78712, USA}

\author[0000-0003-4150-841X]{Elisabeth R. Newton}
\affil{Department of Physics and Astronomy, Dartmouth College, Hanover, NH 03755, USA}




\begin{abstract}

Orbiting a M dwarf 12 pc away, the transiting exoplanet GJ 1132b is a prime target for transmission spectroscopy. With a mass of 1.7 M$_{\oplus}$ and radius of 1.1 R$_{\oplus}$, GJ 1132b's bulk density indicates that this planet is rocky. Yet with an equilibrium temperature of 580K, GJ 1132b may still retain some semblance of an atmosphere. Understanding whether this atmosphere exists and its composition will be vital for understanding how the atmospheres of terrestrial planets orbiting M dwarfs evolve. We observe five transits of GJ 1132b with the Wide Field Camera 3 (WFC3) on the {\it Hubble Space Telescope (HST)}. We find a featureless transmission spectrum from 1.1--1.7$\mu$m, ruling out cloud-free atmospheres with metallicities $<$300$\times$ Solar with $>$4.8$\sigma$ confidence. We combine our WFC3 results with transit depths from {\it TESS} and archival broadband and spectroscopic observations to find a featureless spectrum from 0.7--4.5$\mu$m. GJ 1132b has a high mean molecular weight atmosphere, possesses a high-altitude aerosol layer, or has effectively no atmosphere. Higher precision observations are required to differentiate between these possibilities. We explore the impact of hot and cold starspots on the observed transmission spectrum GJ 1132b, quantifying the amplitude of spot-induced transit depth features. Using a simple Poisson model we estimate spot temperature contrasts, spot covering fractions, and spot sizes for GJ 1132. These limits, and the modeling framework, may be useful for future observations of GJ 1132b or other planets transiting similarly inactive M dwarfs.

\end{abstract}




\section{Introduction} \label{sec:intro}

If our Solar System terrestrial planets and moons are any indicator, rocky exoplanets likely possess a diverse population of atmospheres. From the thick CO$_{2}$-dominated atmosphere of Venus, out to the tenuous N$_{2}$ atmosphere of Pluto, the rocky worlds of our Solar System have undergone significant atmospheric evolution \citep[for a review see][]{solar.system.atmospheres}. Many terrestrial exoplanets have likely experienced similar transformations leading to a variety of atmospheres \citep[e.g.,][]{segura.ozone.2003,grenfell.atmosphere.2007,hu.photochemistry.2012,forget.and.leconte.2014,luger.abiotic.2015,schaefer.gj1132b,kite.atm.loss.2020,grenfell.et.al.2020}. It is also possible that some rocky exoplanets never lost their primary atmospheres. Instead, they continue to maintain a slight H/He envelope consisting of $<$ 0.1\% of their masses initially accreted from the planetary nebula \citep{owen.et.al.2020}. Characterizing these terrestrial atmospheres will be crucial for understanding the formation, evolution, and potential habitability of these rocky worlds.

Of interest for atmospheric characterization are the rocky planets orbiting nearby bright M dwarfs. The smaller stellar radii of these stars translate directly into larger transit depths for Earth-sized planets passing in front of them \citep{nutzman.2008.DesignConsiderationsGroundBased, winn.transit.review}. Moreover, these systems tend to be compact, with the habitable zone generally within 0.2 AU of the star \citep{kopparapu.habitable.zone}. As the probability of a transit is inversely proportional to the semi-major axis, there is a higher likelihood of a M dwarf planet transiting than one orbiting FGK dwarfs.

Therefore, most super-Earth and Earth-sized planets amenable for transmission spectroscopy orbit nearby bright M dwarfs \citep[e.g.][]{bertathompson.et.al.2015,trappist.1.discovery,dittmann.et.al.2017,tess.2019,tess.2019.2,ment.lhs1140c,winters.ltt1445,luque.et.al.2019,ment.toi540,trifonov.et.al.2021}.

One notable planet for atmospheric study is the Earth-sized rocky planet, GJ 1132b. Discovered with the MEarth survey \citep{bertathompson.et.al.2015}, this 1.13 $\pm$ 0.056 R$_{\oplus}$ and 1.66 $\pm$ 0.23 M$_{\oplus}$ rocky planet orbits a nearby inactive M3.5 dwarf \citep{hawley.et.al.1996,dittmann.et.al.2017,Bonfils.et.al.2018}. GJ 1132b receives 19$\times$ more bolometric insolation than Earth, implying a global equilibrium temperature of about 580 K (assuming uniform heat redistribution and bond albedo of 0), making it far too hot to be habitable. GJ 1132b receives more stellar flux than Mercury, but its higher mass and surface gravity lead to the questions: Does GJ 1132b possess an atmosphere? And if so, what is its atmospheric composition? 

GJ 1132b was targeted by ground-based observations seeking answers to these questions \citep{southworth.et.al.2017,diamondlowe.et.al.2018}. In broadband transit depths, \citet{southworth.et.al.2017} marginally detected hints of a low-metallicity atmosphere. This evidence was based primarily on a single anomalously deep z-band transit and required a 20\% larger stellar radius than previously inferred. In contrast, \citet{diamondlowe.et.al.2018} observed a featureless spectrum between 700 and 1040 nm with an average precision of 100 ppm per wavelength bin (equivalent to 1.8 scale heights assuming a H/He mean molecular weight). Based on their higher precision results, they disfavor a cloud-free atmosphere of ${<}10\times$ Solar metallicity by volume. This suggests either a secondary atmosphere with a higher mean molecular weight, high-altitude aerosols, or no atmosphere around GJ 1132b. The broadband 4.5 $\mu$m {\it Spitzer} channel and the MEarth optical 0.7--1.0 $\mu$m observations from \citet{dittmann.et.al.2017} do not by themselves strongly constrain possible atmospheres. They do provide a light-curve-derived stellar density that confirms a stellar radius in agreement with \citet{diamondlowe.et.al.2018} and \citet{bertathompson.et.al.2015}. 

\citet{waalkes.et.al.2019} searched for a deep Ly$\alpha$ ultraviolet transit of GJ 1132b using {\it HST}/STIS. They detected no transits but place a 2$\sigma$ upper limit of an exosphere at 7.3$\times$ the planetary radius. From this, they argue that it is unlikely that GJ 1132b possesses an extended hydrogen envelope created from a leftover primary atmosphere or from the photodissociation of water. Based on the planet's total high-energy irradiation estimated from the stellar Ly$\alpha$ emission, \citet{waalkes.et.al.2019} calculate a mass-loss upper limit of neutral hydrogen to be 0.86 $\times$ 10$^{9}$ g/s, or one Earth ocean every 6 Myr.

Theoretical work shows that any atmosphere of a planet like GJ 1132b has been influenced by the intense X-ray and ultraviolet radiation and activity common to M dwarfs \citep{france.activity}. M dwarfs exhibit more XUV flux relative to bolometric than G dwarfs, making M dwarfs more efficient drivers of atmospheric escape \citep{lopez.and.fortney.2013,owen.and.wu.2017}. Analysis of UV observations of a similar planet host, LHS 3844, back these theoretical arguments while supporting the hypothesis that any atmosphere surrounding GJ 1132b is secondary in nature \citep{diamond.lowe.atmosphere.escape}.

\citet{schaefer.gj1132b} modeled the evolution of GJ 1132b's atmosphere in order to predict potential atmospheric compositions, tracking both atmospheric escape and mantle outgassing. They find that GJ 1132b would have required at least 5\% of its initial mass to be water in order to retain a water-dominated atmosphere today. As this would be unlikely for a planet forming interior to the snow line, \citet{schaefer.gj1132b} argue that the most likely outcome is a tenuous atmosphere dominated by O$_{2}$ created by the photodissociation of H$_{2}$O. However, most of the hydrogen and some of the oxygen atoms produced from this process are lost to space through hydrodynamic drag or absorption into a magma ocean. \citet{schaefer.gj1132b} briefly comment on a CO$_{2}$-dominated atmosphere. They note that adding significant CO$_{2}$ will prolong a surface magma ocean and enhance atmospheric loss. However, future exploration and modeling is required to test this hypothesis.

Here we investigate the existence of an atmosphere around GJ 1132b by observing its transmission spectrum from the \textit{Hubble Space Telescope (HST)} Wide-Field Camera 3 (WFC3) between 1.1 and 1.7 $\mu$m. These observations enable a sensitive search for water-vapor absorption, provide observational constraints on possible atmospheric compositions, and test whether or not GJ 1132b was able to retain a primordial hydrogen-rich atmosphere.  

However, stellar activity, especially unocculted stellar spots and faculae, have the potential to either mimic or mask planetary water features in the WFC3 bandpass \citep{rackham.2018.TransitLightSource,zhang.2018.NearinfraredTransmissionSpectra}. GJ 1132, with a rotation period of 130 days, is not young \citep{newton.gj1132.rotation} but does show variability due to rotating starspots, and \citet{dittmann.et.al.2017} attributed an offset between the MEarth optical and {\it Spitzer} infrared transit depths to unocculted stellar spots. We therefore also use available data to quantify the impact of GJ 1132's stellar activity on the interpretation of planetary transmission spectra observed by both \textit{HST}/WFC3 and future instruments, such as the \textit{James Webb Space Telescope (JWST)} or ground-based giant segmented-mirror telescopes (GSMTs).

We present this work in the following sections. We outline the observations in Section~\ref{sec:observation} and detail the analysis of the broadband and spectroscopic light curves in Section~\ref{sec:analysis}. We present an analysis of the {\it TESS} light curve of GJ 1132 in this section as well, providing additional constraints on the system properties and ruling out the transit of GJ 1132c. Section~\ref{sec:discussion} discusses the transmission spectrum and the possible atmospheric compositions for GJ 1132b. We quantify stellar contamination in the transmission spectrum in Section~\ref{sec:starspot} before concluding in Section~\ref{sec:conclusion}. 

\section{Observations and Data Reduction} \label{sec:observation}

\subsection{Data}
We observed five transits of GJ 1132b with WFC3/IR grism spectroscopy (GO \#14758, PI Berta-Thompson). We used the G141 grism, providing stellar spectra at a resolution of $R = \lambda/\Delta \lambda = 130$ covering approximately 1.1$-$1.7 $\mu$m . Each visit consisted of one orbit used to settle WFC3's optical and detector systematics (which was discarded from analysis), one orbit before transit, one orbit in transit, and one orbit after transit. Phase constraints were set to maximize time in transit between Earth occultations, and the five scheduled visits together provide complete phase coverage of the transit, including both ingress and egress. 

We gathered data with the 256x256 subarray, using the SAMP-SEQ=SPARS10 and NSAMP=15 settings for a total photon-counting exposure time of 88s per exposure. To minimize the overheads associated with detector readout and/or memory buffer dumps, we used a round-trip spatial scan in the cross-dispersion direction. A scan rate of 0.2$\farcs$/second kept the maximum fluence recorded on the detector a safe level of about 24,000 e-/pixel. For each exposure, 72s were lost to readout and scan resets, resulting in an overall photon-counting duty cycle of 55\% for the 22 exposures gathered each orbit between Earth occultations. 

GJ 1132 is near the Galactic plane ($b=8^\circ$) and therefore in a crowded field. This presents a challenge for slitless spectroscopy due to the potential for multiple stars' spectra to overlap. In planning the observations, we picked a combination of ORIENT constraints and scan rate to prevent overlap of any significant background stars with GJ 1132's first-order spectrum during a single non-destructive read.

\subsection{Reduction}
We reduced each of the five WFC3 visits using the publicly available {\tt iraclis} pipeline\footnote{\url{https://github.com/ucl-exoplanets/Iraclis}}. We include a summary of the pipeline's steps here; \citet{iraclis.2016a, iraclis.2016b} provide a full description of the pipeline's methodology. Starting with the raw uncalibrated spatially scanned spectroscopic images, {\tt iraclis} performs the following: a zero-read subtraction, non-linearity correction, dark correction, gain conversion, flat fielding, sky background subtraction, and bad-pixel and cosmic-ray interpolation. {\tt iraclis} calculates potential horizontal and vertical shifts in the scanned spectrum over time and determines the scan length for each image by approximating a Gaussian along the sum of the rows. Based off the direct image of GJ 1132, wavelengths are assigned and the wavelength-dependent flat field is applied. {\tt iraclis} reports times in units of HJD which we converted to BJD$_\mathrm{TDB}$ using the time utilities code of \citet{eastman.et.al.2012}\footnote{\url{http://astroutils.astronomy.ohio-state.edu/time/}}.

We extracted the 1-D spectrum by applying a 166-pixel-wide aperture along the dispersion axis and summing along the cross-dispersion axis for each wavelength. As the average scan length for each image is about 170 pixels, we chose a slightly smaller aperture to minimize potential edge effects of the scan. However, slightly smaller and larger apertures (162 pixels and 172 pixels) had no notable improvements on the results.

Across the five visits, there is less than a 0.2 pixel shift in both the $x-$ and $y-$directions of the scan. A comparison of the residuals, created by subtracting off the best-fit models, to their corresponding $x-$ and $y-$ shifts demonstrated no correlation. We therefore did not apply any correction to these small shifts. We also found no correlation with either the scan length or background values to the residuals.

\begin{figure}
    \centering
    \includegraphics[width=\textwidth]{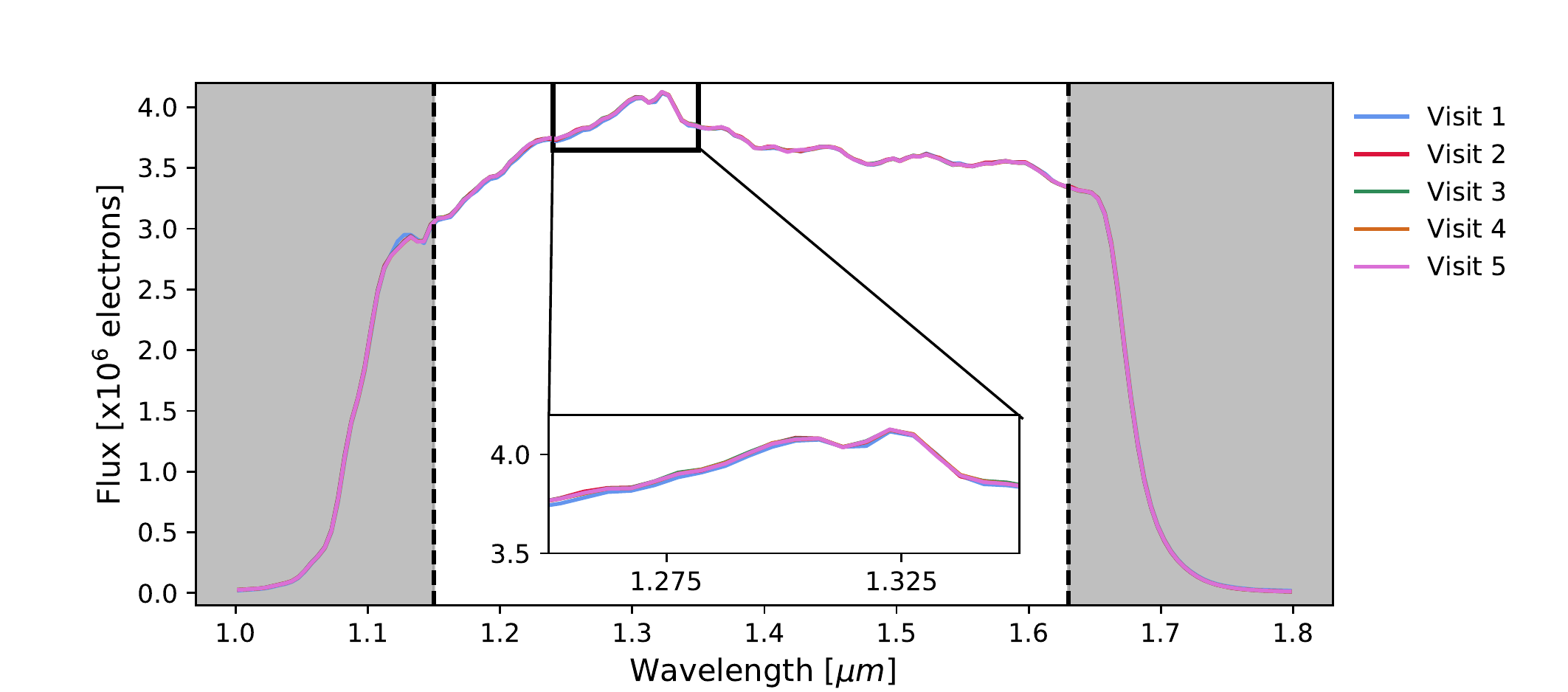}
    \caption{A comparison of the average stellar spectrum extracted for each visit. We find no evidence of significant contamination by other sources as there is very little deviation in flux levels of the star across the five visits. The 1.15$\mu$m and 1.63$\mu$m are marked with dash lines for comparison with Figure~\ref{fig:contamination2}.}
    \label{fig:contamination}
\end{figure}

Due to the crowded stellar field, we checked the direct and dispersed images of each visit for other nearby bright stars in the field of view. A comparison of the average spectrum between visits demonstrated less than a 1\% flux variation across all wavelengths (Figure~\ref{fig:contamination}). We identified a weak 0.5\% background source contaminating the GJ 1132 spectrum for each read, similarly noted in \citet{mugnai.et.al.2021}. This background source contamination imparts a 12 ppm systematic suppression of the overall transit depth, smaller than the 34 ppm wavelength-binned spectroscopic transit depth uncertainties. 

\begin{figure}
    \centering
    \includegraphics[width=\textwidth]{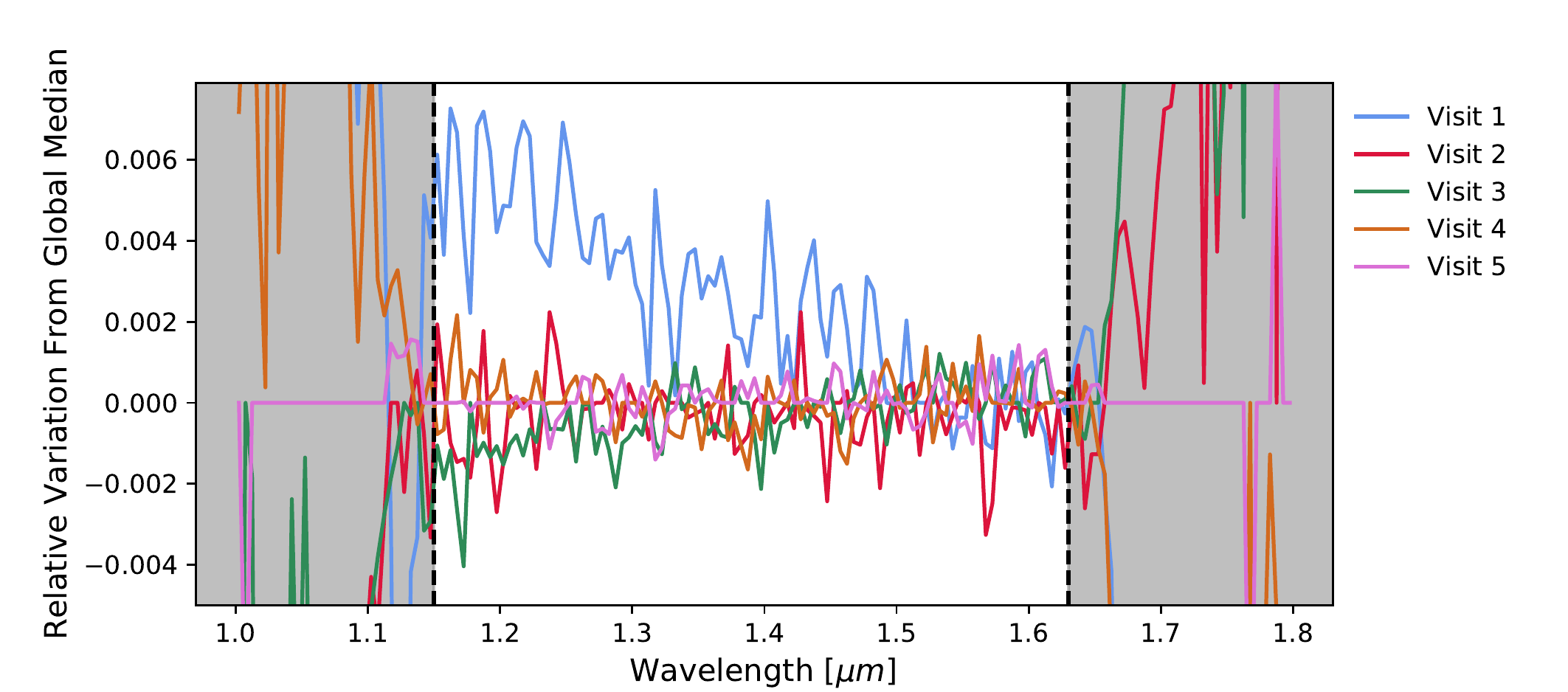}
    \caption{The deviation of each visit from the global medium-combined stellar spectrum created from the combination of all five visits. Visits 2$-$5 shows $<$0.2\% difference across 1.15 and 1.63$\mu$m while visit 1 differs by $<$0.8\% notably at shorter wavelengths.
    Dashed lines flank the high-SNR region.}
    \label{fig:contamination2}
\end{figure}

A comparison of each visit's absolute flux to the global median-combined total flux spectrum shows $<$0.2\% deviation for visits 2$-$5 across 1.15 to 1.63 $\mu$m (Figure~\ref{fig:contamination2}). Visit 1 demonstrates a $<$0.8\% deviation with larger divergence at shorter wavelengths. This difference between the visits is likely a stellar effect. Visits 1 and 2 were observed 158 days apart (1.2$\times$ the stellar rotation rate of 130 days) while visits 2$-$5 were all observed within 64 days of one another (0.5$\times$ the stellar rotation). To mitigate the slight flux variation across all visits, we opted to analyze each visit independently discussed in further detail below.

\section{Analysis}\label{sec:analysis}
We analyzed the WFC3 data both as wavelength-integrated broadband light curves and as binned spectroscopic light curves. We also analyzed archival {\it TESS} data of GJ 1132, enabling both a new optical transit depth for GJ 1132b and a search for transits of the radial-velocity detected planet GJ 1132c \citep{Bonfils.et.al.2018}.

\subsection{HST Broadband Light Curve Analysis} \label{bblc}

From the {\tt iraclis}-reduced spectra, we created broadband light curves for each visit by summing the flux from 1.15 to 1.63 $\mu$m. Uncertainties are a combination of the photon shot noise from the star, sky background, dark current, and read noises added in quadrature. As observed in multiple WFC3 light curves, each orbit demonstrated charge-trapping ramps, with the first orbit demonstrating significantly larger effects \citep[e.g.][]{berta.et.al.2012,kreidberg.et.al.2014,wakeford.et.al.2016,zhou.et.al.2017}. We multiplied the physical charge-trapping instrumental model {\tt RECTE}  \citep{zhou.et.al.2017} with the {\tt BATMAN} transit model \citep{batman.paper} and fitted each visit independently across all parameters (including both transit and instrumental systematic parameters). 

For the {\tt RECTE}  portion of the model, we fitted for the starting number of slow and fast traps ($E_{\rm s,f}$) as well as the number of traps between the orbits for slow and fast traps ($\Delta E_{\rm s,f}$) recommended by \citet{zhou.et.al.2017}. We also discovered that fitting for the total number of slow and fast traps ($E_{\rm s,tot}$ and $E_{\rm f,tot}$) significantly improved the fits for each visit. This is in contrast to \citet{zhou.et.al.2017}, who recommended holding the total number of traps, trapping times, and trapping efficiencies constant. We did find that varying trapping times and efficiencies did not improve our overall fits and held these constant to the values listed in \citet{zhou.et.al.2017}. We included visit-long slopes and flux offsets to the charge-trapping model with these parameters dependent on the scanning direction. While \citet{zhou.et.al.2017} note that one benefit of {\tt RECTE}  is the inclusion of the first orbit, we discovered that removing this orbit improved the overall fits for the three remaining orbits. \citet{guo.et.al.2020} also found this to be true in their WFC3 analysis of HD 97658b. {\tt RECTE}  is designed to approximate the charge trapping occurring per pixel. We therefore assumed an average per-pixel flux value by dividing the total flux in an image by the number of pixels contained in the aperture before fitting the systematics with {\tt RECTE}.

For the {\tt BATMAN}  transit portion of the model, we performed two independent versions of the transit fit: (1) varying $a/R_{s}$ and inclination and (2) holding these two parameters constant to the values determined by \citet{dittmann.et.al.2017} from high-cadence {\it Spitzer} data. For both versions, we fitted for the mid-transit time and $R_{p}/R_{s}$. Each visit only sampled the ingress or egress of the transit. We therefore held the quadratic limb-darkening coefficients constant, which were determined by LDTK \citep{ldtk.2015} in the WFC3 bandpass using GJ 1132's stellar parameters: $\log{g}=5.049$ \citep{dittmann.et.al.2017}, $Z=-0.12$ \citep{bertathompson.et.al.2015}. We used the Gaia-determined stellar effective temperature of 3630 K \citep{gaia.et.al.2018b} in calculating the limb-darkening coefficients. This $T_{\rm eff}$ differs from the more accurate 3270 K \citet{bertathompson.et.al.2015} we adopt throughout the rest of this work; we re-ran one transit with the coefficients for this cooler temperature and found the parameter results for broadband and spectroscopic fits to be negligible.

The period remained fixed to that determined by \citet{dittmann.et.al.2017}. We held the eccentricity constant at 0, which is supported by previous RV measurements from \citet{bertathompson.et.al.2015} and \citet{Bonfils.et.al.2018}. We also integrated the transit model over the WFC3 total exposure time of 103 seconds; accounting for this finite integration time is particularly important because it is comparable to the ingress/egress time \citep[see][]{kipping.2010.Binningsinningmorphological}.

Our final model, combining both {\tt RECTE}  and {\tt BATMAN},  for the flux $F_{\rm for,rev}(t)$ as a function of both the forward and reverse scan direction and time $t$ is therefore:
\begin{equation}
\begin{split}
F_{\rm for,rev}(t) =& (F_{\rm for,rev} + m_{\rm for,rev}t)\times \\
& \mathrm{RECTE}(E_{s,f}, \Delta E_{s,f}, E_{s_{tot},f_{tot}})\times \\
& \mathrm{BATMAN}(R_{\rm p}/R_{\rm s},t_{\rm 0},(a/R_{s})^{ver}, i^{ver}).
\end{split}
\label{eq:rectebatman}
\end{equation}

F$_{\rm for,rev}$ and m$_{\rm for,rev}$ represent the initial offset and visit long slope for the forward and reverse scanning directions respectively. Parameters denoted with $\{s,f\}$ in the {\tt RECTE} model refer to slow and fast trap parameters respectively, and parameters with `ver' depend on the transit version in question. These parameters, combined with an error scaling parameter, lead to 15 and 13 fitted parameters for the first and second versions respectively.

We used the MCMC fitter {\tt emcee} \citep{mcmc.paper} to fit the above model to each of the five visits separately. An error scaling term (f$_{scale}$) was also included, which was multiplied to the flux uncertainties in order to allow the MCMC to inflate the uncertainties to achieve $\chi^{2}_{r} = 1$ and propagate appropriately into the parameter distributions. Uniform priors were assumed for all parameters. We assigned 100 walkers with 15,000 steps. After an initial run, we check for any outliers that fell $>$5$\sigma$ away from the median absolute deviation (MAD). These points were flagged and removed, and the fit was performed again. This accounted for two outlier points in visit 2 and one outlier point in visit 3. We binned the residuals from this fit and found a 1/$\sqrt{N}$ Gaussian slope demonstrating that we properly isolated and removed systematic noise. 

We decided to perform the two versions noted above as the transit duration of each visit is largely unconstrained. By varying $a/R_{s}$ and inclination for the first fit, we determined that all visits have less than a 2$\sigma$ deviation from those values quoted in \citet{dittmann.et.al.2017}. Therefore, we adopted \citet{dittmann.et.al.2017} values for these parameters ($a/R_{s}$: 16.54 $i$: 88.68) for our second fit. By holding $a/R_{s}$ and inclination constant, we determined a more precise $R_{p}/R_{s}$ and mid-transit time, though there is little deviation in the overall value between the two versions for both parameters. Table~\ref{tab:planet_params} lists our best fit planetary parameters from both versions for the five {\it HST} visits. The second version parameters are used for all further analyses including the transmission spectrum analysis in Section~\ref{sec:spectra_analysis}. We included the parameters determined from the {\it TESS} light curve (Section~\ref{sec:tess_analysis}) to Table~\ref{tab:planet_params} for comparison. The error scaling parameter ($f_{scale}$) fitted for during the MCMC routine and $\chi_{r}^{2}$ are listed as goodness-of-fit measurements. 

We calculated the stellar density using the $a/R_{s}$ from each broadband transit and found that those of all but visit 2 were within 1$\sigma$ of the 29.6$\pm$6.0\,g/cm$^3$ reported in \citet{bertathompson.et.al.2015}. Visit 2 has the least number of points defining the ingress or egress, making it a challenge to constrain the transit duration (and therefore stellar density). \citet{southworth.et.al.2017} measured a lower stellar density of 15.4$^{+4.8}_{-3.4}$ g/cm$^3$, which gave them larger stellar and planetary radii; we find that all of our visits except for 1 and 2 have a $>$2$\sigma$ deviation from this value. We therefore opt to use the stellar and planetary radii derived in \citet{bertathompson.et.al.2015} and \citet{Bonfils.et.al.2018}.

\begin{figure}
    \centering
    \includegraphics[width=\textwidth]{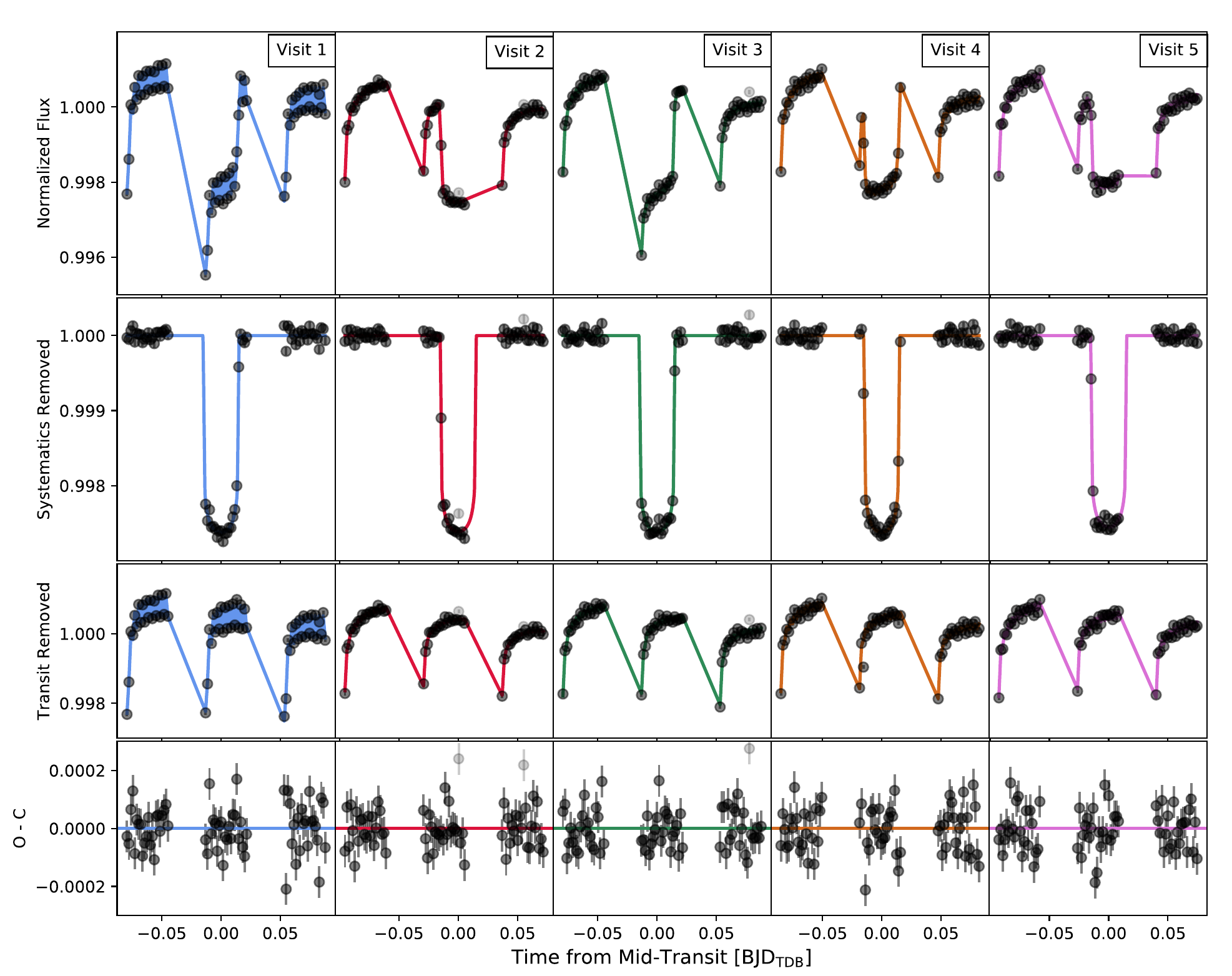}
    \caption{Broadband light curves for each WFC3 transit of GJ 1132b, with each visit plotted left to right, including the raw light curves with transit and ramp-like systematics ({\it first row}), the transits with the {\tt RECTE}  systematic model divided out ({\it second row}), the systematics with the {\tt BATMAN}  transit model divided out ({\it third row}), and the final residuals ({\it fourth row}). Outlying points are noted as light gray points.}
    \label{fig:four_panel}
\end{figure}

We plot the five broadband light curves with the best-fit models from the second version in Figure~\ref{fig:four_panel}. From top-down each panel shows: the normalized flux, the transit with the {\tt RECTE}  model removed, the systematics with the {\tt BATMAN}  transit model removed, and the residuals. Visit 1 has the largest deviation in flux between the forward and reverse scans, giving the ramp an up-and-down motion that is not as apparent in the other four visits. The origin of this difference is unclear, though the systematics model captures this effect. We do not plot the best fit from the first version, as the difference between the fits is slight and not apparent when plotted against each other.

\begin{table}[]
\scriptsize
\caption{Best-fit parameters for GJ 1132b across the five {\it HST} visits and the folded {\it TESS} light curve.}
\label{tab:planet_params}
\begin{tabular}{lrrrrrr}
\hline
\multicolumn{1}{l|}{}                               & \multicolumn{1}{c|}{Visit 1}                     & \multicolumn{1}{c|}{Visit 2}                     & \multicolumn{1}{c|}{Visit 3}                     & \multicolumn{1}{c|}{Visit 4}                     & \multicolumn{1}{c|}{Visit 5} & \multicolumn{1}{c}{TESS} \\ \hline
\multicolumn{6}{c}{-------version 1: $a/R_{s}$ and $i$ free parameters-------}                                                                                                                                                                                                                           \\ \hline
\multicolumn{1}{l|}{$R_{p}/R_{s}$}                  & \multicolumn{1}{r|}{0.0502 $\pm$ 0.0007}         & \multicolumn{1}{r|}{0.0494 $\pm$ 0.0007}         & \multicolumn{1}{r|}{0.0489 $\pm$ 0.0004}         & \multicolumn{1}{r|}{0.0491 $\pm$ 0.0004}         & \multicolumn{1}{r|}{0.0485 $\pm$ 0.0005} & 0.0519 $\pm$ 0.0014        \\
\hline
\multicolumn{1}{l|}{Mid-Transit Time ($T_{0}$)} & \multicolumn{1}{r|}{862.19219} & \multicolumn{1}{r|}{1020.20402} & \multicolumn{1}{r|}{1077.21000} & \multicolumn{1}{r|}{1080.46869} & \multicolumn{1}{r|}{1083.72525} &  1544.70993\\


\multicolumn{1}{l|}{(-2457000 BJD)} & \multicolumn{1}{r|}{$\pm$ 0.00051} & \multicolumn{1}{r|}{$\pm$ 0.00327} & \multicolumn{1}{r|}{$\pm$ 0.00054} & \multicolumn{1}{r|}{$\pm$ 0.00007} & \multicolumn{1}{r|}{$\pm$ 0.00090} &  $\pm$ 0.00074\\
\hline
\multicolumn{1}{l|}{$a/R_{s}$}                      & \multicolumn{1}{r|}{13.61$^{+3.15}_{-0.79}$}     & \multicolumn{1}{r|}{11.02$^{+2.32}_{-1.60}$}     & \multicolumn{1}{r|}{16.65$^{+0.68}_{-1.27}$}     & \multicolumn{1}{r|}{16.47$^{+0.77}_{-1.30}$}     & \multicolumn{1}{r|}{17.59$^{+1.69}_{-2.55}$} & 15.59$^{+0.89}_{-2.18}$     \\
\hline
\multicolumn{1}{l|}{$i$ (deg)}                      & \multicolumn{1}{r|}{$87.34^{+2.10}_{-0.40}$}     & \multicolumn{1}{r|}{$88.22^{+1.27}_{-1.81}$}     & \multicolumn{1}{r|}{89.26 $\pm$ 0.81}            & \multicolumn{1}{r|}{88.91 $\pm$ 0.74}            & \multicolumn{1}{r|}{89.19$^{+0.56}_{-1.16}$} & 88.77$^{+0.87}_{-1.28}$     \\
\hline
\multicolumn{1}{l|}{$f_{scale}$}                    & \multicolumn{1}{r|}{1.29 $\pm$ 0.11}             & \multicolumn{1}{r|}{1.17 $\pm$ 0.10}             & \multicolumn{1}{r|}{1.24 $\pm$ 0.12}             & \multicolumn{1}{r|}{1.19 $\pm$ 0.10}             & \multicolumn{1}{r|}{1.21 $\pm$ 0.11} & 0.87 $\pm$ 0.01             \\
\hline
\multicolumn{1}{l|}{$\chi^{2}_{r}$ (dof)}           & \multicolumn{1}{r|}{2.31 (51)}                   & \multicolumn{1}{r|}{1.99 (49)}                   & \multicolumn{1}{r|}{2.57 (50)}                   & \multicolumn{1}{r|}{2.15 (51)}                   & \multicolumn{1}{r|}{2.24 (51)} & 0.80 (1971)                   \\ \hline
\multicolumn{6}{c}{-------version 2: fixed $a/R_{s}$: 16.54 $i$: 88.68-------}                                                                                                                                                                                                                           \\ \hline
\multicolumn{1}{l|}{$R_{p}/R_{s}$}                  & \multicolumn{1}{r|}{0.0493 $\pm$ 0.0003}         & \multicolumn{1}{r|}{0.0490 $\pm$ 0.0002}         & \multicolumn{1}{r|}{0.0492 $\pm$ 0.0002}         & \multicolumn{1}{r|}{0.0493 $\pm$ 0.0003}         & \multicolumn{1}{r|}{0.0486 $\pm$ 0.0003} & 0.0515 $\pm$ 0.0012         \\
\hline

\multicolumn{1}{l|}{Mid-Transit Time ($T_{0}$)} & \multicolumn{1}{r|}{862.19257} & \multicolumn{1}{r|}{1020.19808} & \multicolumn{1}{r|}{1077.21151} & \multicolumn{1}{r|}{1080.46896} & \multicolumn{1}{r|}{1083.72617} & 1544.71421\\

\multicolumn{1}{l|}{(-2457000 BJD)} & \multicolumn{1}{r|}{$\pm$ 0.00006} & \multicolumn{1}{r|}{$\pm$ 0.00004} & \multicolumn{1}{r|}{$\pm$ 0.00005} & \multicolumn{1}{r|}{$\pm$ 0.00004} & \multicolumn{1}{r|}{$\pm$ 0.00005} & $\pm$ 0.00023\\
\hline
\multicolumn{1}{l|}{$f_{scale}$}                    & \multicolumn{1}{r|}{1.31 $\pm$ 0.10}             & \multicolumn{1}{r|}{1.21 $\pm$ 0.11}             & \multicolumn{1}{r|}{1.11 $\pm$ 0.10}             & \multicolumn{1}{r|}{1.32 $\pm$ 0.12}             & \multicolumn{1}{r|}{1.23 $\pm$ 0.11} & 0.87 $\pm$ 0.01             \\
\hline
\multicolumn{1}{l|}{$\chi^{2}_{r}$ (dof)}           & \multicolumn{1}{r|}{2.42 (53)}                   & \multicolumn{1}{r|}{1.47 (51)}                   & \multicolumn{1}{r|}{1.79 (52)}                   & \multicolumn{1}{r|}{2.49 (53)}                   & \multicolumn{1}{r|}{2.15 (53)} & 0.76 (1974)                  
\end{tabular}
\end{table}

\subsection{HST Spectroscopic Light Curve Analysis}\label{sec:spectra_analysis}

We created 22 spectroscopic light curves by summing the flux between 1.15 to 1.63 $\mu$m into 5-pixel bins (approximately 20 nm each). We analyzed these light curves using two different methods: fitting each light curve using {\tt RECTE}  and using the broadband divide-white method described in \citet{kreidberg.et.al.2014}.

\paragraph{RECTE model} We again approximated a per-pixel average by dividing the total summed flux in the spectroscopic channel by the number of pixels in the smaller binned aperture. Similar to the broadband light curve analysis, we did not include the first orbit. We did include the number of slow and fast traps as fitted parameters. LDTK was used to determine the quadratic limb darkening parameters for each wavelength bin, which were held constant throughout each individual fit. $a/R_{s}$ and inclination were held constant to the \citet{dittmann.et.al.2017} values. The mid-transit time was held constant to that obtained from the second-version WFC3 broadband fit. This left $R_{p}/R_{s}$ as the only free transit model parameter, along with the 10 systematic parameters from the {\tt RECTE}  model and an error scaling term.

\begin{figure}
    \centering
    \includegraphics[width=.6\textwidth]{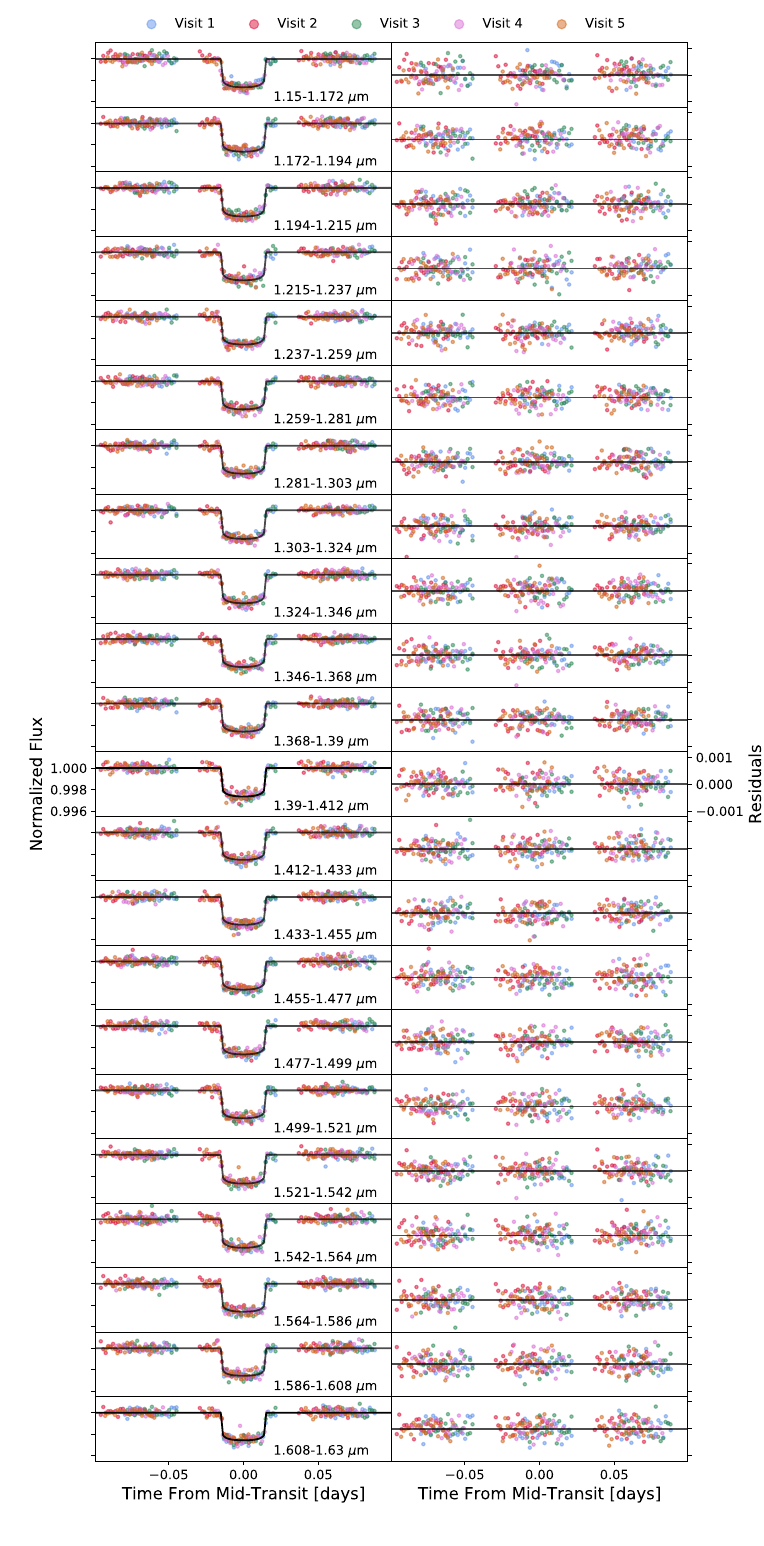}
    \caption{Each of the 22 spectroscopic light curves plotted with the top to bottom being smallest to largest wavelength bins. Point colors correspond to one of the five visits. Light curves are shown with systematics removed and with the transit model created from the weighted-average transit depth of the five visits ({\it left}), and residuals from this shared model are shown ({\it right}).}
    \label{fig:spectrocurves}
\end{figure}

Each spectroscopic light curve for each visit was modeled individually with the same {\tt emcee} MCMC process as the broadband curves. The $R_{p}/R_{s}$ for each visit and the corresponding wavelength bins are noted in Table~\ref{tab:wavedepth}. Inverse-variance weighted averages calculated from the five visits are listed in the last column of the table. Each spectroscopic light curve for all five visits is plotted in Figure~\ref{fig:spectrocurves} with the systematics-removed transit curve plotted on the left and the residuals for each bin to the right. We find little deviation in the {\tt RECTE}  model parameters from bin to bin, highlighting their mostly wavelength-independent nature. 

\paragraph{Divide-White Method} We checked our {\tt RECTE} analysis by applying the divide-white method to the spectroscopic light curves \citep{kreidberg.et.al.2014}. We created systematic residuals (third row in Figure~\ref{fig:four_panel}) by dividing the broadband transit model from the broadband data. By assuming these systematic residuals are not wavelength-dependent, we divide these residuals from each spectroscopic light curve. We find that while the residuals captured the ramp-like effect of the data, they do not necessarily remove the difference in the forward and backward scanning offset. We therefore fit for two separate offsets, dependent on the scanning direction, as well as the transit depth and an error-scaling term. Again, we performed the same {\tt emcee} MCMC routine as the broadband and {\tt RECTE}  analyses discussed above.

\begin{figure}
    \centering
    \includegraphics[width=\textwidth]{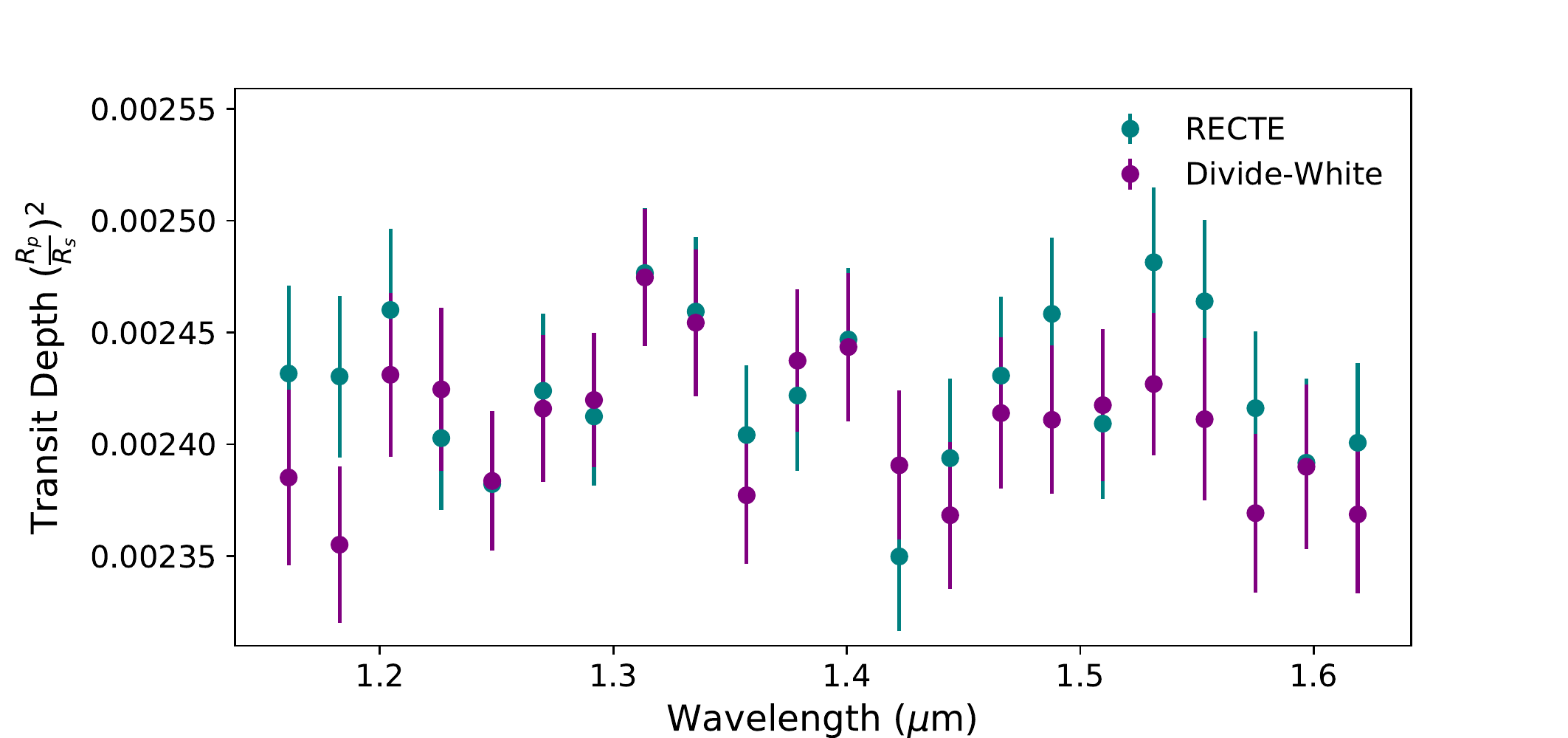}
    \caption{A transit depth comparison between the {\tt RECTE}  (green) and divide-white (purple) methods for each wavelength bin. Overall, transit depths vary less than 2$\sigma$ for each bin (except for the second bin) with an average discrepancy of less than 1$\sigma$.}
    \label{fig:method_compare}
\end{figure}

A comparison of the transit depths between the {\tt RECTE} and divide-white methods demonstrated discrepancies that were $<$2$\sigma$ for all but the second wavelength bins (Figure~\ref{fig:method_compare}). The {\tt RECTE} method always produced slightly larger error bars on the transit depths due to the larger number of free parameters. We opted to use the {\tt RECTE} transit depths (R$_{p}$/R$_{s}$)$^{2}$ for the rest of the analysis and discussion in this paper as this model is based on physical properties of the detector. 

\begin{table}[]
\scriptsize
\caption{The R$_{p}$/R$_{s}$ for each wavelength bin across each visit. We include the inverse-variance weighted average of the visits in the last column. To reproduce the transit depths plotted throughout this paper, we use (R$_{p}$/R$_{s}$)$^{2}$ as the transit depth and 2(R$_{p}$/R$_{s}$)$\sigma_{R_{p}/R_{s}}$ to calculate the transit depth uncertainty.}
\label{tab:wavedepth}
\begin{tabular}{c|r|r|r|r|r|r}
\multicolumn{1}{c|}{Wavelength {[}$\mu m${]}} & \multicolumn{1}{c|}{Visit 1} & \multicolumn{1}{c|}{Visit 2} & \multicolumn{1}{c|}{Visit 3} & \multicolumn{1}{c|}{Visit 4} & \multicolumn{1}{c|}{Visit 5} & \multicolumn{1}{c}{\textbf{Weighted Average}} \\ \hline
1.150-1.172                                    & 0.0480 $\pm$ 0.0009          & 0.0507 $\pm$ 0.0011          & 0.0480 $\pm$ 0.0009          & 0.0480 $\pm$ 0.0009          & 0.0480 $\pm$ 0.0009          & \textbf{0.0493 $\pm$ 0.0004}                 \\
1.172-1.194                                    & 0.0505 $\pm$ 0.0009          & 0.0489 $\pm$ 0.0008          & 0.0505 $\pm$ 0.0009          & 0.0505 $\pm$ 0.0009          & 0.0505 $\pm$ 0.0009          & \textbf{0.0493 $\pm$ 0.0004}                 \\
1.194-1.215                                    & 0.0495 $\pm$ 0.0008          & 0.0493 $\pm$ 0.0008          & 0.0495 $\pm$ 0.0008          & 0.0495 $\pm$ 0.0008          & 0.0495 $\pm$ 0.0008          & \textbf{0.0496 $\pm$ 0.0004}                 \\
1.215-1.237                                    & 0.0490 $\pm$ 0.0007          & 0.0488 $\pm$ 0.0007          & 0.0490 $\pm$ 0.0007          & 0.0490 $\pm$ 0.0007          & 0.0490 $\pm$ 0.0007          & \textbf{0.0490 $\pm$ 0.0003}                 \\
1.237-1.259                                    & 0.0495 $\pm$ 0.0007          & 0.0491 $\pm$ 0.0007          & 0.0495 $\pm$ 0.0007          & 0.0495 $\pm$ 0.0007          & 0.0495 $\pm$ 0.0007          & \textbf{0.0488 $\pm$ 0.0003}                 \\
1.259-1.281                                    & 0.0492 $\pm$ 0.0008          & 0.0479 $\pm$ 0.0008          & 0.0492 $\pm$ 0.0008          & 0.0492 $\pm$ 0.0008          & 0.0492 $\pm$ 0.0008          & \textbf{0.0492 $\pm$ 0.0004}                 \\
1.281-1.303                                    & 0.0496 $\pm$ 0.0006          & 0.0496 $\pm$ 0.0008          & 0.0496 $\pm$ 0.0006          & 0.0496 $\pm$ 0.0006          & 0.0496 $\pm$ 0.0006          & \textbf{0.0491 $\pm$ 0.0003}                 \\
1.303-1.324                                    & 0.0487 $\pm$ 0.0006          & 0.0502 $\pm$ 0.0007          & 0.0487 $\pm$ 0.0006          & 0.0487 $\pm$ 0.0006          & 0.0487 $\pm$ 0.0006          & \textbf{0.0498 $\pm$ 0.0003}                 \\
1.324-1.346                                    & 0.0494 $\pm$ 0.0007          & 0.0509 $\pm$ 0.0007          & 0.0494 $\pm$ 0.0007          & 0.0494 $\pm$ 0.0007          & 0.0494 $\pm$ 0.0007          & \textbf{0.0496 $\pm$ 0.0003}                 \\
1.346-1.368                                    & 0.0488 $\pm$ 0.0006          & 0.0482 $\pm$ 0.0008          & 0.0488 $\pm$ 0.0006          & 0.0488 $\pm$ 0.0006          & 0.0488 $\pm$ 0.0006          & \textbf{0.0490 $\pm$ 0.0003}                 \\
1.368-1.390                                    & 0.0482 $\pm$ 0.0008          & 0.0501 $\pm$ 0.0009          & 0.0482 $\pm$ 0.0008          & 0.0482 $\pm$ 0.0008          & 0.0482 $\pm$ 0.0008          & \textbf{0.0492 $\pm$ 0.0003}                 \\
1.390-1.412                                    & 0.0510 $\pm$ 0.0007          & 0.0496 $\pm$ 0.0010          & 0.0510 $\pm$ 0.0007          & 0.0510 $\pm$ 0.0007          & 0.0510 $\pm$ 0.0007          & \textbf{0.0495 $\pm$ 0.0003}                 \\
1.412-1.433                                    & 0.0502 $\pm$ 0.0009          & 0.0481 $\pm$ 0.0007          & 0.0502 $\pm$ 0.0009          & 0.0502 $\pm$ 0.0009          & 0.0502 $\pm$ 0.0009          & \textbf{0.0485 $\pm$ 0.0003}                 \\
1.433-1.455                                    & 0.0491 $\pm$ 0.0008          & 0.0487 $\pm$ 0.0008          & 0.0491 $\pm$ 0.0008          & 0.0491 $\pm$ 0.0008          & 0.0491 $\pm$ 0.0008          & \textbf{0.0489 $\pm$ 0.0004}                 \\
1.455-1.477                                    & 0.0505 $\pm$ 0.0009          & 0.0481 $\pm$ 0.0009          & 0.0505 $\pm$ 0.0009          & 0.0505 $\pm$ 0.0009          & 0.0505 $\pm$ 0.0009          & \textbf{0.0493 $\pm$ 0.0004}                 \\
1.477-1.499                                    & 0.0507 $\pm$ 0.0007          & 0.0497 $\pm$ 0.0007          & 0.0507 $\pm$ 0.0007          & 0.0507 $\pm$ 0.0007          & 0.0507 $\pm$ 0.0007          & \textbf{0.0496 $\pm$ 0.0003}                 \\
1.499-1.521                                    & 0.0497 $\pm$ 0.0008          & 0.0500 $\pm$ 0.0009          & 0.0497 $\pm$ 0.0008          & 0.0497 $\pm$ 0.0008          & 0.0497 $\pm$ 0.0008          & \textbf{0.0491 $\pm$ 0.0003}                 \\
1.521-1.542                                    & 0.0503 $\pm$ 0.0007          & 0.0497 $\pm$ 0.0007          & 0.0503 $\pm$ 0.0007          & 0.0503 $\pm$ 0.0007          & 0.0503 $\pm$ 0.0007          & \textbf{0.0498 $\pm$ 0.0003}                 \\
1.542-1.564                                    & 0.0493 $\pm$ 0.0008          & 0.0493 $\pm$ 0.0009          & 0.0493 $\pm$ 0.0008          & 0.0493 $\pm$ 0.0008          & 0.0493 $\pm$ 0.0008          & \textbf{0.0496 $\pm$ 0.0004}                 \\
1.564-1.586                                    & 0.0497 $\pm$ 0.0008          & 0.0473 $\pm$ 0.0009          & 0.0497 $\pm$ 0.0008          & 0.0497 $\pm$ 0.0008          & 0.0497 $\pm$ 0.0008          & \textbf{0.0492 $\pm$ 0.0004}                 \\
1.586-1.608                                    & 0.0496 $\pm$ 0.0008          & 0.0489 $\pm$ 0.0009          & 0.0496 $\pm$ 0.0008          & 0.0496 $\pm$ 0.0008          & 0.0496 $\pm$ 0.0008          & \textbf{0.0489 $\pm$ 0.0004}                 \\
1.608-1.630                                    & 0.0499 $\pm$ 0.0010          & 0.0486 $\pm$ 0.0008          & 0.0499 $\pm$ 0.0010          & 0.0499 $\pm$ 0.0010          & 0.0499 $\pm$ 0.0010          & \textbf{0.0490 $\pm$ 0.0004}                
\end{tabular}

\end{table}

\subsection{TESS Light Curve Analysis of GJ 1132b}\label{sec:tess_analysis}

\begin{figure}
    \centering
    \includegraphics[width=\textwidth]{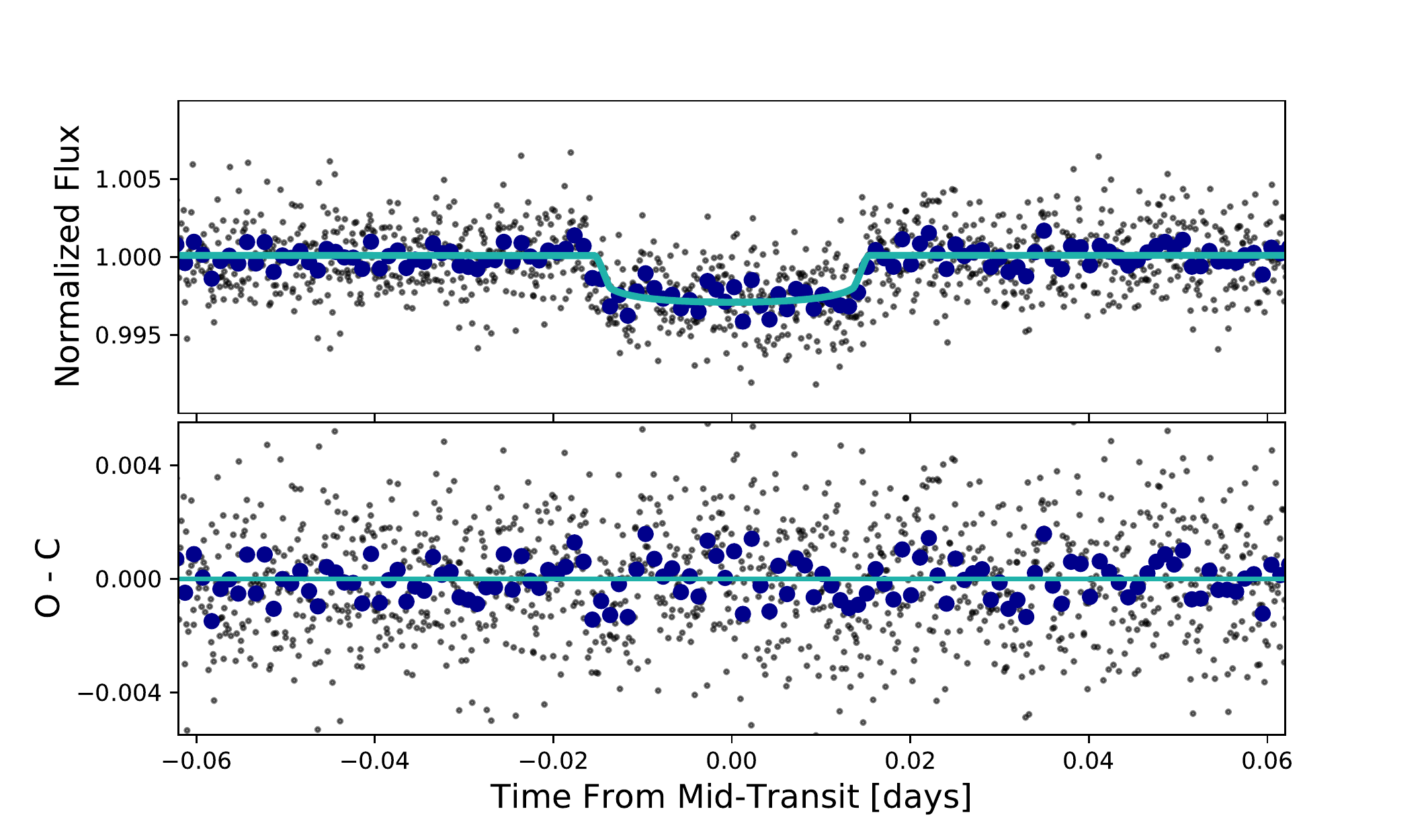}
    \caption{{\it TESS} folded light curve of GJ 1132b's transit. Top panel includes the detrended light curve points (gray), binned points for clarity (blue), and the best-fit model in light blue. Bottom panel shows the residuals from this fit.}
    \label{fig:tesstransit}
\end{figure}

GJ 1132 was observed by {\it TESS} during Sectors 9 and 10. In Sector 10, the star appeared on a noisier section of the CCD, demonstrating larger instrumental variation. For this analysis, we therefore opted to use only the Sector 9 data, including a total of 14 individual transits. We used the python package {\tt lightkurve} \citep{lightkurve.cite} to download and analyze the two-minute-cadence observations. Starting from the PDC light curve, we detrended the flux by masking out all GJ 1132b transits and applying a median moving boxcar filter with a baseline of one day ($32\times$ the transit duration). As with the broadband WFC3 transits, we performed the same two versions of fits as above (varying $a/R_{s}$ and inclination and holding them constant) to all transits simultaneously. We include a baseline of 135 minutes (3$\times$ the transit duration) on either side of each transit. For both versions, we fitted for $R_{p}/R_{s}$, a transit ephemeris, quadratic limb-darkening coefficients, and a general offset. We assumed uniform priors on all parameters except for the limb darkening coefficients which were assigned Gaussian priors based on values calculated by LDTK in the {\it TESS} bandpass \citep{ldtk.2015}. We include the error scaling parameter. Table~\ref{tab:planet_params} lists the best-fit parameters for both versions, and Figure~\ref{fig:tesstransit} plots the best-fit model atop the folded data.  We found no significant difference between the two versions; the fitted values of $a/R_{s}$ and inclination were consistent with those from {\it Spitzer} \citep{dittmann.et.al.2017}.

We also analyzed individual {\it TESS} transits, again applying the two versions. Using the $a/R_{s}$ and inclination values for each transit, we calculated the transit duration for each epoch. We see no evidence for transit duration variations or transit depth variations in the {\it TESS} Sector 9 data. A brief inspection of Sector 10 showed large variations in each due to significant instrumental noise in the data even after the median detrending, supporting our decision to remove this sector from our analysis.

\begin{figure}
    \centering
    \includegraphics[width=\textwidth]{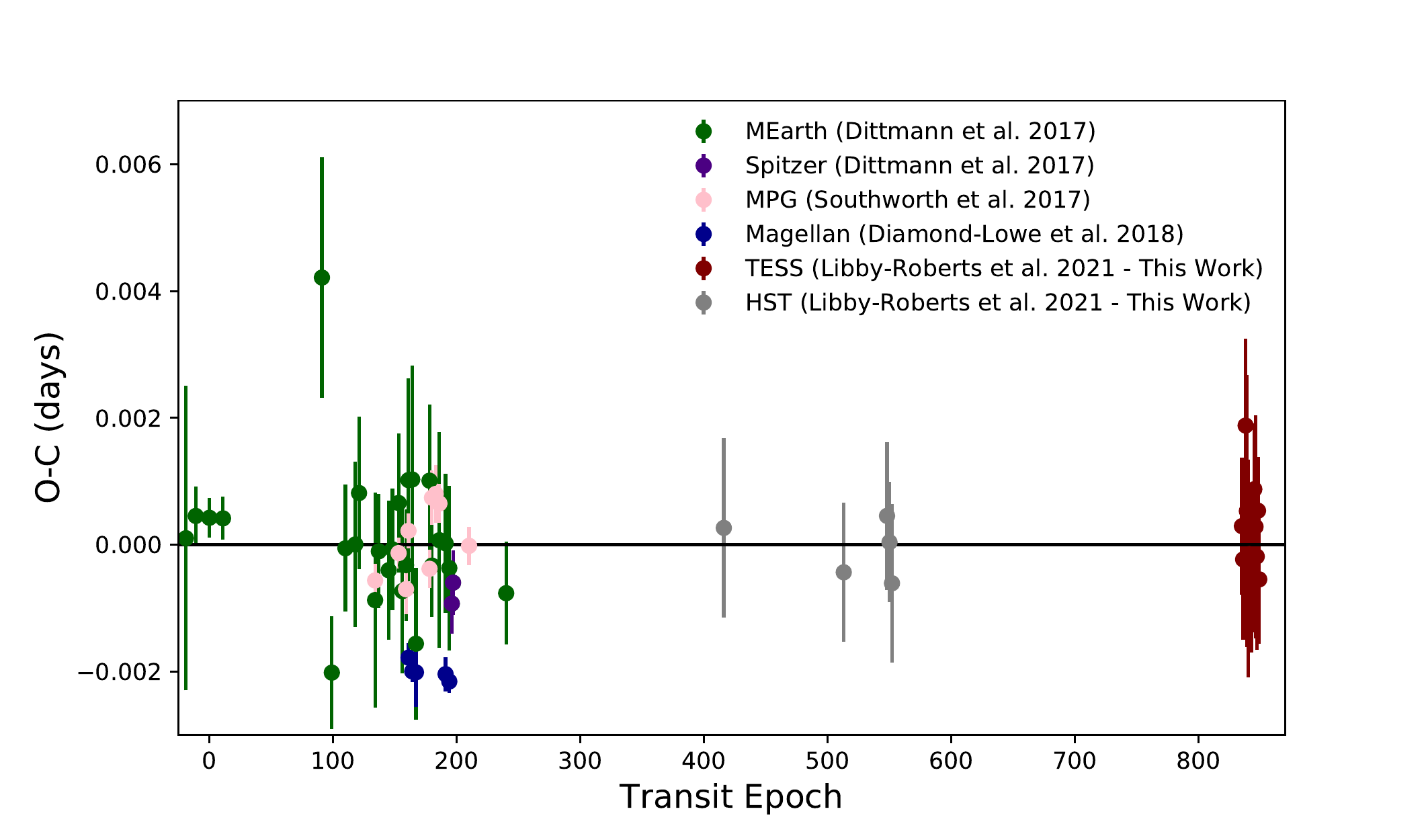}
    \caption{Searching for TTVs in the mid-transit times of GJ 1132b yielded scatter around a flat line. Points are color-coded by telescope, and the {\it HST} uncertainties are inflated to match the 103-second exposure time of one observation point.}
    \label{fig:ttv_plot}
\end{figure}

We measured the Sector 9 mid-transit times for each transit, holding $a/R_{s}$ and inclination constant. We combined these times to the broadband mid-transit times for each {\it HST} visit as well as the published times from MEarth \citep{bertathompson.et.al.2015, dittmann.et.al.2017}, {\it Spitzer} \citep{dittmann.et.al.2017}, and MPG telescope \citep{southworth.et.al.2017}. \citet{diamondlowe.et.al.2018} noted that their mid-transit times for GJ 1132b were consistently 2 minutes early for all four of their observed transits and were unable to determine the cause of this difference. We also observed this discrepancy and do not include their times for our analysis. The uncertainties on the mid-transit times determined from the WFC3 observations are on the order of four seconds, less than the exposure time of one data point. To more cautiously capture this uncertainty, we inflate the {\it HST} mid-transit uncertainties by a factor of 25 in order to match the 103 second exposure time of one point. Using the BJD$_{\mathrm{TDB}}$ mid-transit times for each epoch, we fit a straight line through these points providing an updated period of 1.6289299 $\pm$ 5x10$^{-7}$ days and a transit ephemeris of 2457184.55747 $\pm$ 0.00012 BJD$_{\mathrm{TDB}}$. Table~\ref{tab:times} reports the mid-transit times for {\it HST} broadband curves with the inflated uncertainties and times from Sector 9 of TESS. A search for transit timing variations (TTVs) yielded a non-detection with the residuals matching a flat line with $\chi^{2}_{\rm r}=1.17$ for 53 degrees of freedom (Figure~\ref{fig:ttv_plot}).

\begin{table}[]
\begin{tabular}{crc}
\multicolumn{1}{c}{Transit Epoch} & \multicolumn{1}{c}{$T_{0}$ (-2457000 BJD$_{\mathrm{TDB}}$)} & Telescope \\ \hline
416                               & 862.19257 $\pm$ 0.00142\footnote{Uncertainties have been inflated 25$\times$ the 1$\sigma$ value}                      & HST        \\
513                               & 1020.19808 $\pm$ 0.00109$^{\mathrm{a}}$                     & HST        \\
548                               & 1077.21151 $\pm$ 0.00117$^{\mathrm{a}}$                     & HST        \\
550                               & 1080.46896 $\pm$ 0.00095$^{\mathrm{a}}$                     & HST        \\
552                               & 1083.72617 $\pm$ 0.00125$^{\mathrm{a}}$                     & HST        \\
835                               & 1544.71423 $\pm$ 0.00108                           & TESS       \\
836                               & 1546.34264 $\pm$ 0.00126                           & TESS       \\
837                               & 1547.98548 $\pm$ 0.00408                           & TESS       \\
838                               & 1549.60261 $\pm$ 0.00136                           & TESS       \\
839                               & 1551.23019 $\pm$ 0.00214                           & TESS       \\
840                               & 1552.85822 $\pm$ 0.00171                           & TESS       \\
841                               & 1554.48737 $\pm$ 0.00095                           & TESS       \\
843                               & 1557.74480 $\pm$ 0.00112                           & TESS       \\
844                               & 1559.37409 $\pm$ 0.00116                           & TESS       \\
845                               & 1561.00412 $\pm$ 0.00104                           & TESS       \\
846                               & 1562.63245 $\pm$ 0.00176                           & TESS       \\
847                               & 1564.26092 $\pm$ 0.00148                           & TESS       \\
848                               & 1565.89057 $\pm$ 0.00085                           & TESS       \\
849                               & 1567.51841 $\pm$ 0.00101                           & TESS      
\end{tabular}
\caption{Individual {\it HST} and {\it TESS} Sector 9 Mid-Transit Times.}
\label{tab:times}
\end{table}

\subsection{TESS Light Curve Analysis of GJ 1132c}\label{sec:1132c}

Discovered with radial velocities by \citet{Bonfils.et.al.2018}, GJ 1132c has not yet been completely vetted for the possibility of transits.  \citet{Bonfils.et.al.2018} estimate a 1\% transit probability. No obvious transits were seen in MEarth or {\it Spitzer} data, but they lacked complete and continuous coverage of the possible times of transit. We use {\it TESS's} continual coverage to search for the possibility of a GJ 1132c transit. We predict two possible transit events of GJ 1132c to occur in each Sector. As with our GJ 1132b {\it TESS} analysis, we limited our transit search to Sector 9. We masked out all GJ 1132b transits in the median boxcar detrended data. We assumed Gaussian priors on the $a/R_{s}$, mid-transit time and period parameters using values determined by \citet{Bonfils.et.al.2018} and set uniform priors on $R_{p}/R_{s}$ and the inclination. We allow for grazing transits by letting the impact parameter to vary from 0 to 1$+R_{p}/R_{s}$. We also allow for the possibility of negative transit depths. The quadratic limb darkening coefficients were held constant to those values determined from the GJ 1132b {\it TESS} fits.

\begin{figure}
    \centering
    \includegraphics[width=\textwidth]{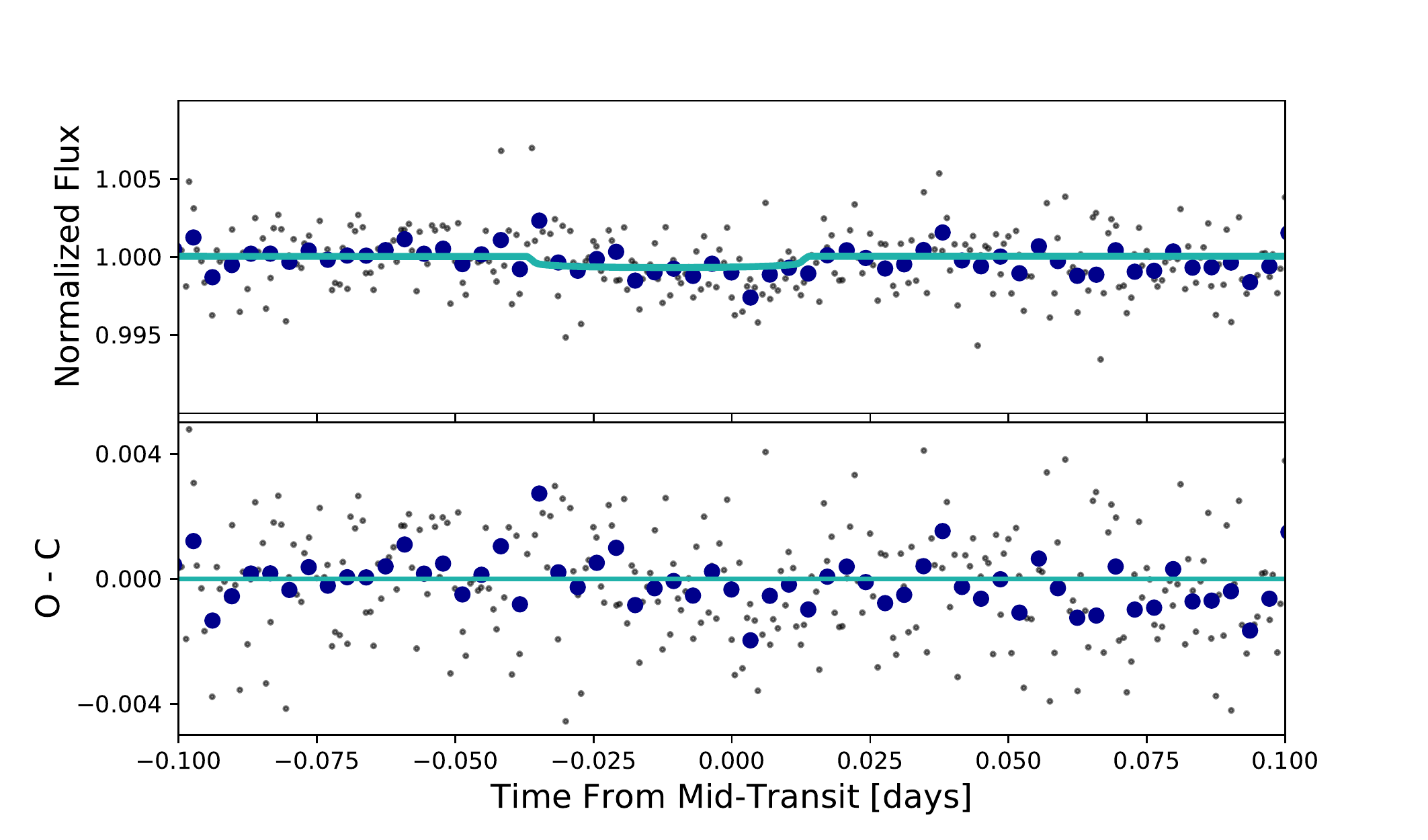}
    \caption{The {\it TESS} Sector 9 light curve of GJ 1132, folded assuming the best-fit period (8.922 $\pm$ 0.009 days) from our {\it TESS} analysis of GJ 1132c, with the best-fit maximum likelihood model plotted in light blue. We find no significant evidence that GJ 1132c transits.}
    \label{fig:gj1132c}
\end{figure}

We performed an MCMC fit using 100 walkers and 15,000 steps with 5,000 of those removed for burn-in. We determined an $R_{p}/R_{s}$ of 0.024$^{+0.011}_{-0.046}$, a result consistent with zero. A visual inspection of the data shows no indication of a obvious transit (Figure~\ref{fig:gj1132c}). From the posterior, we place a 3$\sigma$ upper limit on $R_{p}/R_{s}$ at 0.081, marginalized over all other parameters. This corresponds to a planet radius of 1.84 R$_{\oplus}$. With a minimum mass of 2.64 M$_{\oplus}$ \citep{Bonfils.et.al.2018}, it is possible that this planet possesses a radius smaller than what we can detect with two {\it TESS} transits. More data from the {\it TESS} extended mission will be required to completely rule out a GJ 1132c transit. We also add an additional caveat that a box least squares (BLS) algorithm would provide a more rigorous search of the data, especially as the RV parameters on the period and expected mid-transit time are not well-constrained. However, we calculate that from GJ 1132c's a/R$_{s}$ \citep{Bonfils.et.al.2018}, its orbit would need to be inclined to at least 89.88 degrees (1.2 degrees from GJ 1132b's orbit) in order to transit. While this is close to co-planar with its transiting neighbor, the tight constraint on GJ 1132c's inclination for transits to occur makes it more likely that it is inclined such that it does not transit.

\section{Transmission Spectrum and Comparison to Atmospheric Models}\label{sec:discussion}

\subsection{Transmission Spectrum}\label{subsec:spectrum}

\subsubsection{{\it HST}/WFC3 Transmission Spectrum}

\begin{figure}
    \centering
    \includegraphics[width=\textwidth]{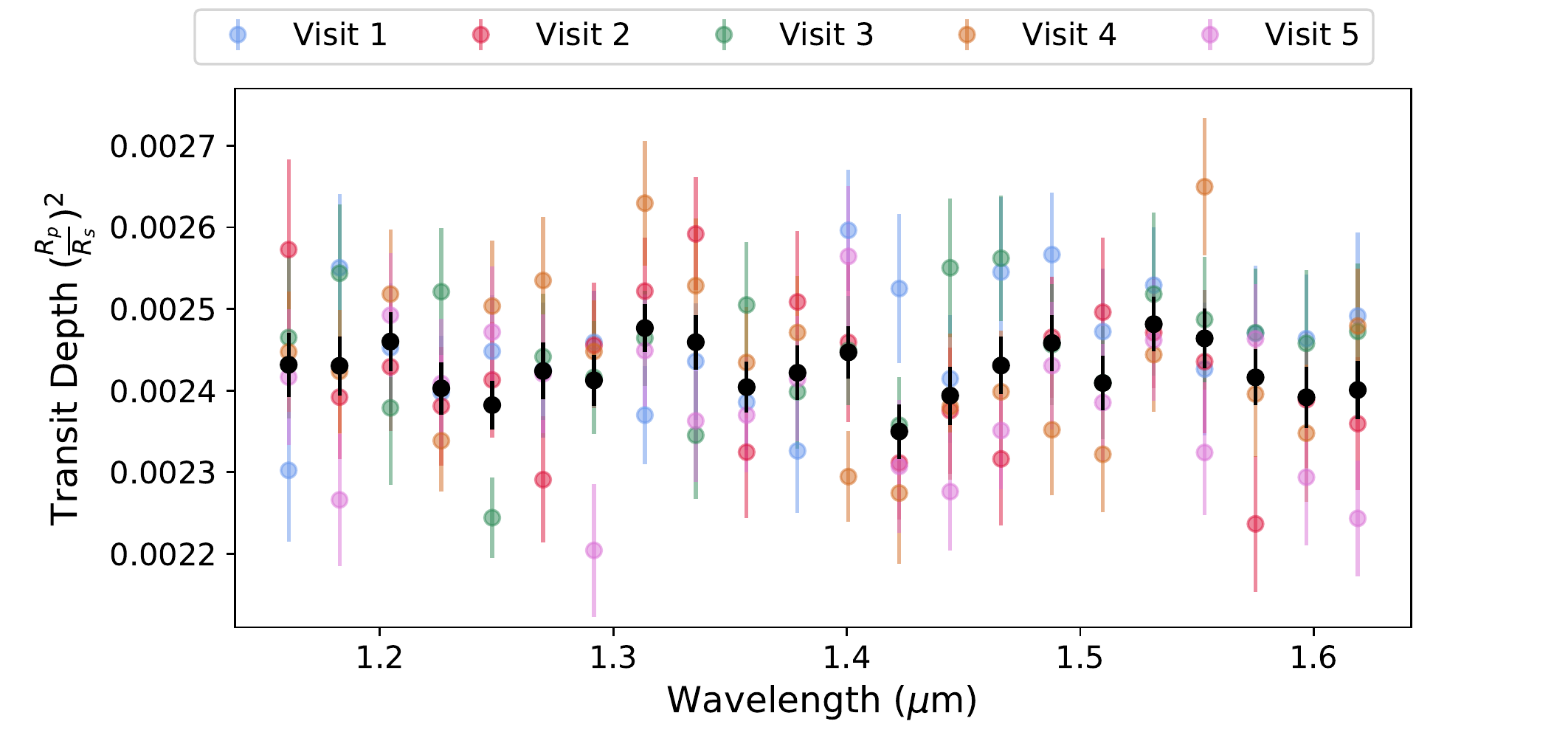}
    \caption{A comparison of the transit depths from each visit in their respective wavelength bins. The weighted-mean transit depths between the five visits are included in black. All bins demonstrated $<$2$\sigma$ spread, except for wavelengths 1.25 and 1.40 $\mu$m. Removing the outlying visits in these bins had no appreciable effect on the final spectrum.}
    \label{fig:visit_compare}
\end{figure}

Before combining the five visits into a weighted-average spectrum, we checked that the fitted transit depths in each bin are drawn from the same distribution and there are no significant outliers. We determined that every individual transit depth fell within 2$\sigma$ of the weighted-average depth in their respective bin, except for the 1.25 and 1.40 $\mu$m bins (Figure~\ref{fig:visit_compare}). Both bins showed a 2.2$\sigma$ spread with visits 3 and 4 yielding the outlying points respectively. A comparison between the average transit depths with and without these visits yielded no difference on our final results. We therefore keep all five visits when calculating the average transit depth for each wavelength bin.

\subsubsection{Comparison to Other {\it HST} Analyses}

\begin{figure}
    \centering
    \includegraphics[width=\textwidth]{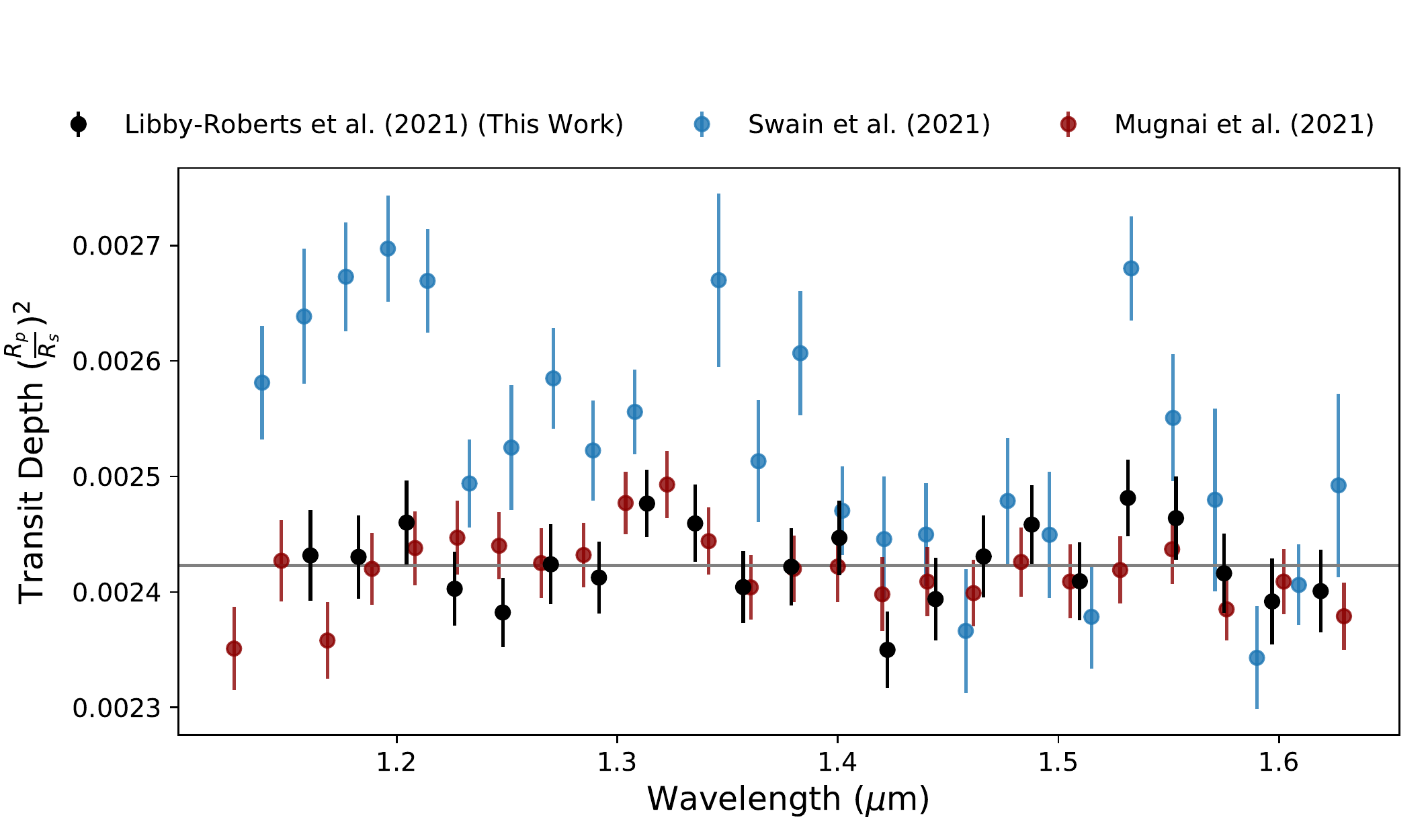}
    \caption{A comparison between the average transit depths in \citet{swain.et.al.2021} (blue points), in \citet{mugnai.et.al.2021} (red points), and this work (black points). All utilize the same {\it HST}/WFC3 data, though \citet{swain.et.al.2021} determine very different transit depths, notably at shorter wavelengths. Transit depths derived in this work and in \citet{mugnai.et.al.2021} are statistically the same with similar uncertainties as well.}
    \label{fig:swain_compare}
\end{figure}

Recently two other papers have published analyses of the same {\it HST}/WFC3 dataset \citep{swain.et.al.2021,mugnai.et.al.2021}, and they reach two different conclusions. \citet{swain.et.al.2021} saw large-amplitude features and inferred a H/He-rich atmosphere with a Rayleigh scattering slope, HCN, and CH$_{4}$ absorption indicating the presence of haze. \citet{mugnai.et.al.2021} measured a featureless spectrum with transit depths and uncertainties similar to ours.  Figure~\ref{fig:swain_compare} compares the transmission spectrum results of those two analyses to the WFC3 transmission spectrum that we measure here. Our inferred transit depths disagree strongly with those from \citet{swain.et.al.2021} and closely agree with those from \citet{mugnai.et.al.2021}. All three of these works use different approaches to analyze and fit the same underlying data. Both our work and \citet{mugnai.et.al.2021} utilized the {\tt iraclis} pipeline. We particularly emphasize the use of a flexible physically-motivated model for instrumental systematics and marginalizing over its parameters at every stage of our fits. We are encouraged by the strong level of consistency we see across our five independent visits. Moreover, for our work to match the results from \citet{swain.et.al.2021}, it would require our analysis to introduce systematic errors that precisely offset the real underlying wave-dependent variation. As this is very unlikely, we conclude that the transmission spectrum of GJ 1132b is effectively flat to 34 ppm precision across the WFC3/G141 bandpass.

\subsubsection{Comparison to Other Archival Data}
\label{sec:archival}

\begin{figure}
    \centering
    \includegraphics[width=\textwidth]{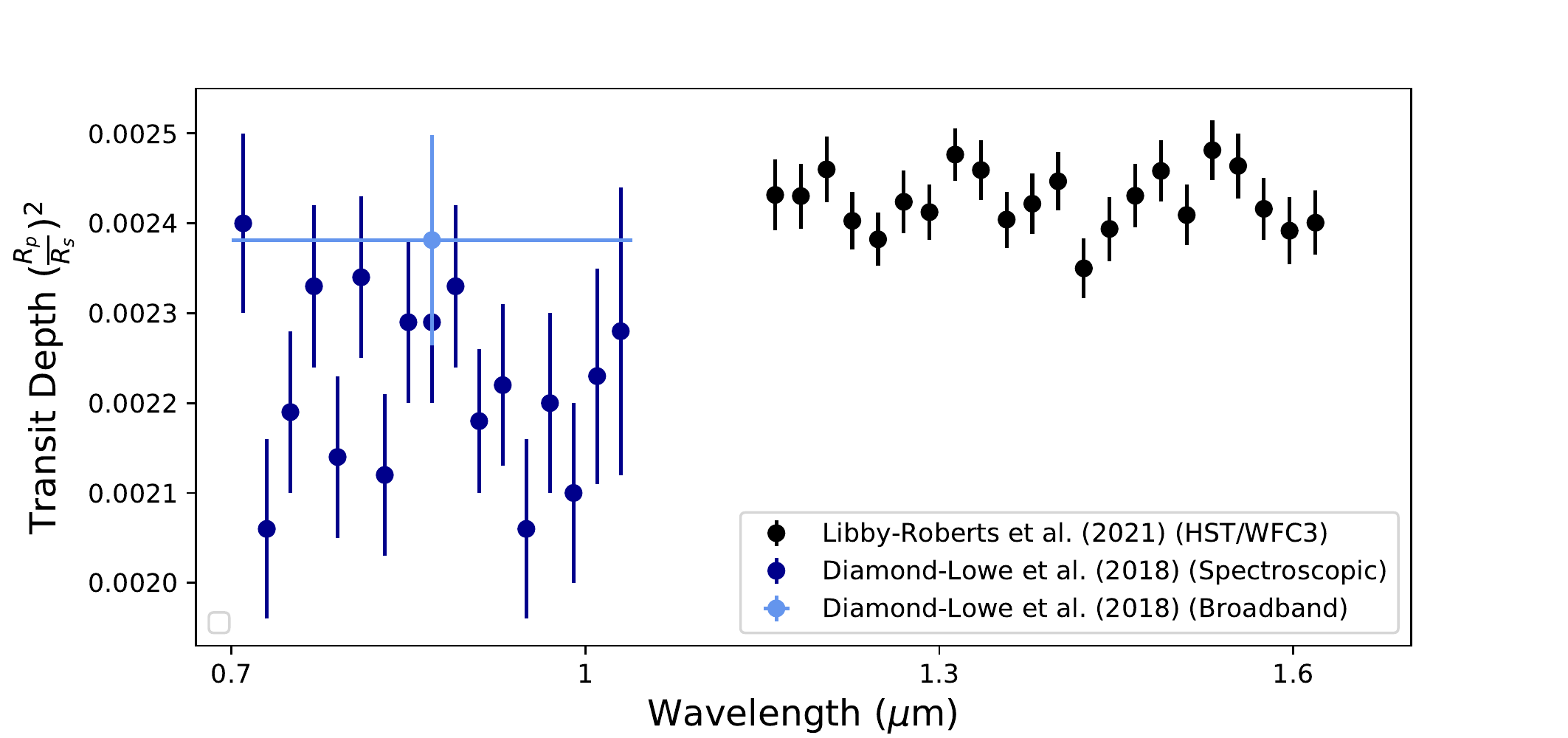}
    \caption{A comparison between the WFC3 transit depths (black points) and the spectroscopic (dark blue) and broadband (light blue) transit depths from \citet{diamondlowe.et.al.2018}. While the spectroscopic depths appear smaller than the WFC3 values, the broadband depth is statistically the same.}
    \label{fig:diamond-lowe-compare}
\end{figure}

We compare the WFC3 transit depths with archival data along with the broadband {\it TESS} point derived in Section~\ref{sec:tess_analysis}. For each data set, we provide a brief summary below:

\begin{itemize}
    \item \citet{dittmann.et.al.2017} observed 21 transits of GJ 1132b with MEarth and 2 transits with the 4.5 $\mu$m {\it Spitzer} channel. They found a {\it Spitzer} transit depth 3$\sigma$ larger than the optical MEarth depth; a feature they attributed to unocculted starspots. We re-analyze the MEarth data (see below) and find a transit depth similar to that of the \citet{dittmann.et.al.2017} {\it Spitzer} depth. 
    
    \item We re-analyzed the 21 MEarth transits reported in \citet{dittmann.et.al.2017} by performing a simple Levenberg-Marquardt fitting routine. We fitted for a common R$_{p}$/R$_{s}$, a/R$_{s}$, and impact parameter (b) across all 21 transits. Airmass coefficients for each transit (21 total coefficients) were also fitted for as well as 132 magnitude zero points (one for each combination of night, telescope and meridian side) \citep{irwin.eclipsing.binary}. We held GJ 1132b's period and transit ephemeris constant to values in Table 3 of \citet{dittmann.et.al.2017}. The constant limb darkening parameters were derived from stellar models \citep{stellar.models} for a star with T$_{\mathrm{eff}}$: 3300K, and log(g): 5.0 and translated to the MEarth bandpass using the same method described in \citet{irwin.eclipsing.binary}. Parameter uncertainties were scaled to account for overdispersion globally rather than individually per night. This re-analysis yielded a deeper transit depth of 0.00246 $\pm$ 0.00011, a 3$\sigma$ difference than the shallower transit found in \citet{dittmann.et.al.2017}.  It is also statistically the same ($<$1$\sigma$) as both the {\it Spitzer} 4.5 $\mu$m and the WFC3 depths. As such, this updated depth indicates no significant stellar contamination between the shorter optical and longer infrared wavelengths.

    \item \citet{southworth.et.al.2017} observed 9 broadband transits with the GROND instrument on the 2.2m MPG telescope. They measure larger transit depths in both the $z$ and $K$ bandpasses, which suggest large H$_{2}$O or CH$_{4}$ absorption in the atmosphere of GJ 1132b. Due to their large transit depth uncertainties and disagreement with other more precise data sets, we do not include the \citet{southworth.et.al.2017} broadband points in any of our analyses.
    
    \item \citet{diamondlowe.et.al.2018} used the LDSS3C spectrograph on the Magellan Clay telescope to observe the optical spectrum of GJ 1132b from 0.7 to 1.04 $\mu$m. They found a featureless spectrum and rule out a 1$\times$ Solar metallicity atmosphere with 3.6$\sigma$ confidence. However, comparing their transit depths to the WFC3 spectrum depicts a slight offset, with their optical transit depths appearing smaller than those in the infrared. A close inspection shows that the broadband transit depth reported in \citet{diamondlowe.et.al.2018} is deeper than any of their spectroscopic depths while also being statistically similar to the depths derived from the WFC3 observations (Figure~\ref{fig:diamond-lowe-compare}). We hypothesize that this difference is due to instrumental and systematic effects and not astrophysical in nature. We add a fitted offset of 0.000205 to the spectroscopic transit depths such that they best fit the various metallicity models which we discuss below.

    \item \citet{swain.et.al.2021} used these {\it HST}/WFC3 data to infer a Rayleigh scattering slope and a possible HCN and CH$_{4}$ absorption feature. We do not observe this in our analysis, and we use our featureless WFC3 spectrum for the comparison with the other data sets.
    
    \item \citet{mugnai.et.al.2021} also used these {\it HST}/WFC3 data but determined a featureless transmission spectrum in agreement with our transit depths. Again, we opt to use our WFC3 depths for the rest of the discussion.
    
\end{itemize}

\subsection{Comparison to Atmospheric Models}

\subsubsection{Cloud-Free Atmospheric Models}

\begin{figure}
    \centering
    \includegraphics[width=\textwidth]{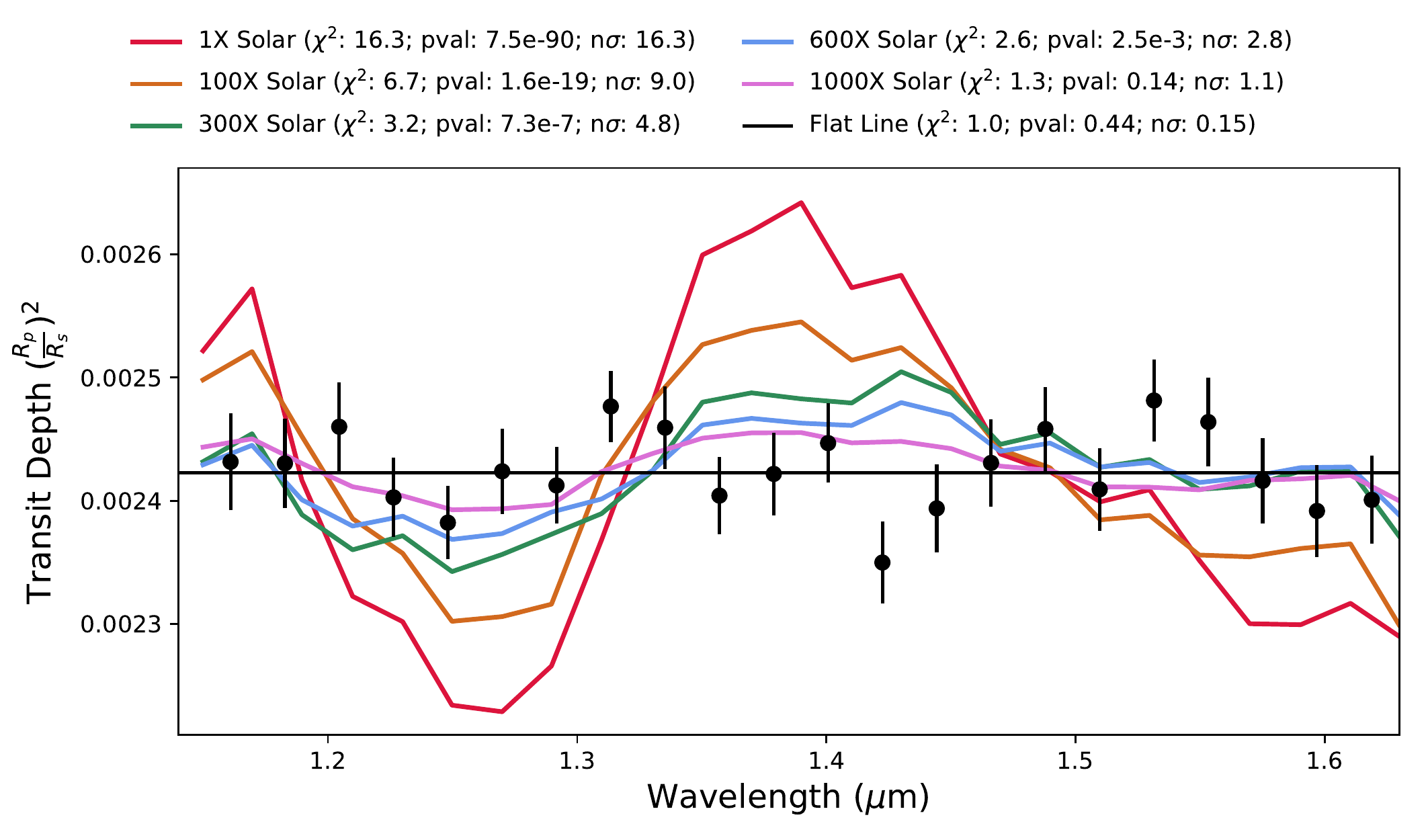}
    \caption{The WFC3 transmission spectrum of GJ 1132b, averaged over the five visits of WFC3, plotted against cloud-free atmospheric models of varying metallicities for comparison. For each model, we provided the $\chi^{2}_{r}$ and the $p$-value evidence against the model. We translated the $p$-value to the number of $\sigma$ (n$\sigma$) confidence that we can reject this model. The spectrum is most consistent with a flat line indicating a high mean molecular weight atmosphere or no atmosphere. We could easily detect an atmosphere with a metallicity less than 300$\times$ Solar by volume (or mean molecular weights less than 8.9 amu).}
    \label{fig:hst_metal}
\end{figure}

We present the WFC3 weighted-average transit depths for GJ 1132b in Figure~\ref{fig:hst_metal}. The planet's transmission spectrum is best explained by a flat or featureless spectrum with a $\chi^{2}_{r}$ of 1.02 with 21 degrees of freedom. The average transit depth uncertainty is 34 ppm per 20 nm wavelength bin. On GJ 1132b, an H/He-dominated atmosphere with a mean molecular weight of $\mu=2.2$ would have a scale height of 172 km; the WFC3 transit depth uncertainties correspond to an effective resolution of 0.29 of those scale heights. In turn, this precision corresponds to 1 scale height for an atmosphere with $\mu = 7.6$. We expect absorption features in the atmosphere to possess sizes of several scale heights. From this precision and the featureless spectrum, we can confidently rule out atmospheres with mean molecular weights less than $\mu = 7.6$. 

\begin{figure}
    \centering
    \includegraphics[width=\textwidth]{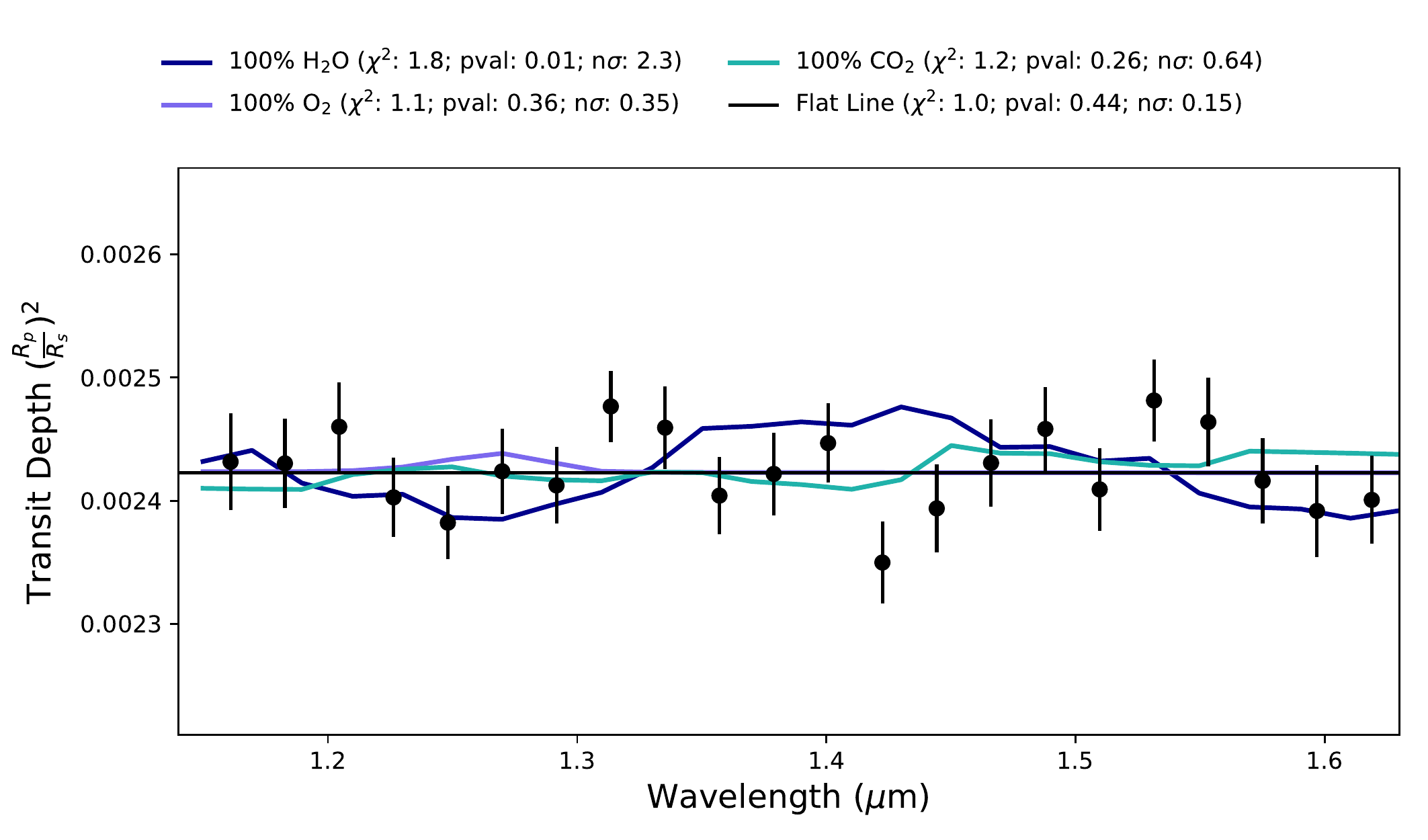}
    \caption{Same as Figure~\ref{fig:hst_metal} except with 100\% molecular composition atmospheric models for comparison. We are unable to rule out any of these atmospheric models as their higher mean molecular weights produce smaller transit depth variations than the precision of the data points.}
    \label{fig:hst_mole}
\end{figure}

For interpreting these transit depths, we generated model transmission spectra using {\tt Exo-Transmit} \citep{exotransmit.cite} for atmospheric compositions of 1$\times$, 100$\times$, 300$\times$, 600$\times$, and 1000$\times$ Solar metallicities by volume (Figure~\ref{fig:hst_metal}) as well as atmospheres composed of 100\%  H$_{2}$O, CO$_{2}$, and O$_{2}$ (Figure~\ref{fig:hst_mole}). These were all cloud-free and excluded any additional Rayleigh scattering beyond the gaseous species present in the atmosphere. Model pressure-temperature (P-T) profiles used the custom double-gray profiles (one opacity for short-wave radiation and another opacity for long-wave radiation) from \citet{diamondlowe.et.al.2018}. See \citet{diamondlowe.et.al.2018} for additional details on the calculations of these profiles. Stellar mass and radius were adopted from \citet{Bonfils.et.al.2018}. As the 1 bar pressure radius is unknown, we use GJ 1132b's planet radius to generate {\tt Exo-Transmit} models for each metallicity before re-scaling them by a multiplicative factor less than 1 in order to best fit the {\it HST} transmission spectrum \citep{exotransmit.cite,diamondlowe.et.al.2018}.

From the WFC3 observations, we confidently rule out atmospheres with metallicities $<$300$\times$ Solar by volume at $>$4.8$\sigma$ significance. This corresponds an atmospheric mean molecular weight $>$8.9 amu (if an atmospheric exists). For comparison, a 1$\times$ Solar atmosphere has a mean molecular weight of 2.2 amu, 100\% H$_{2}$O is 18 amu, O$_{2}$ is 32 amu, and CO$_{2}$ is 44 amu. From these values, we calculate that 8.9 amu could represent an atmosphere mixed with 44\% H$_{2}$O and 56\% H/He by volume. Model atmospheres of 600$\times$ Solar and 1000$\times$ Solar metallicities fit the WFC3 spectrum with $\chi^{2}_{r}$ of 2.6 ($p$-value of 0.0025) and 1.3 ($p$-value of 0.14) respectively.

However, given GJ 1132b's equilibrium temperature of 580 K \citep{Bonfils.et.al.2018}, it is more likely that this planet possesses either a  CO$_{2}$-dominated atmosphere similar to that of Venus, a tenuous O$_{2}$ atmosphere as discussed in \citet{schaefer.gj1132b}, or no atmosphere at all. When comparing atmospheric models of 100\% H$_{2}$O, O$_{2}$, and CO$_{2}$ with the WFC3 data, we calculate $\chi^{2}_{r}$ of 1.8 ($p$-value of 0.01), 1.1 ($p$-value of 0.34), and 1.2 ($p$-value of 0.26) respectively. From these statistics and as illustrated in Figure~\ref{fig:hst_mole}, we cannot rule out any of these scenarios with the current WFC3 precision. We note, however, that GJ 1132b's equilibrium temperature makes it very unlikely that this planet possesses a H$_{2}$O-rich atmosphere.

\subsubsection{Models With Clouds and Hazes}

Featureless exoplanet transmission spectra can be attributed to two different causes: a higher mean molecular weight atmosphere (and thus smaller scale height) or a high-altitude aerosol layer muting or blocking absorption features. While GJ 1132b's temperature and terrestrial nature suggest a higher mean molecular weight atmosphere, the WFC3 observations alone cannot rule out the possibility of a H/He dominated atmosphere combined with condensate clouds such as ZnS or KCl \citep{morley.et.al.2013} or organic photochemical hazes. We therefore determine the altitude required for an aerosol layer to mute all features in the WFC3 transmission spectrum. Assuming a 1$\times$ Solar composition, we place aerosol layers at increasing altitudes until we obtain a spectrum that agrees with the data  \citep[$p$-value of 0.001; process described in detail in][]{libbyroberts.et.al.2020}. We find that this layer must lie at pressure levels $<$0.4 mbars in order to mute the features of a H/He dominated atmosphere. For comparison, \citet{kreidberg.et.al.2014} determine the cloud layer on the sub-Neptune GJ 1214b lies at pressure levels $<$0.01 mbars assuming a low mean molecular weight atmosphere.

We test the possible existence of an aerosol layer above a 1$\times$ Solar atmosphere by calculating GJ 1132b's mass loss rate for a H/He primordial envelope. \citet{owen.et.al.2020} show that low-mass planets similar to GJ 1132b can accrete around 1\% of their masses in H/He gas during formation. With a current mass of 1.66 M$_{\oplus}$, this would translate to a starting H/He envelope mass of 0.0166 M$_{\oplus}$. \citet{waalkes.et.al.2019} approximated a neutral hydrogen mass-loss rate of 0.86 $\times$10$^{9}$ g/s. Assuming no secondary sources of hydrogen and a constant mass-loss rate, GJ 1132b would lose such an atmosphere in 3.7 Gyr. \citet{bertathompson.et.al.2015} approximate this system to be $>$5 Gyr. Moreover, the stellar flux from GJ 1132 should have been higher in the past which would have increased this rate. Therefore, it is very unlikely that GJ 1132b could possess a present-day cloudy primordial H/He-rich atmosphere. 

\subsubsection{GJ 1132b's Atmospheric Composition Determined From Archival Data}

\begin{figure}
    \centering
    \includegraphics[width=\textwidth]{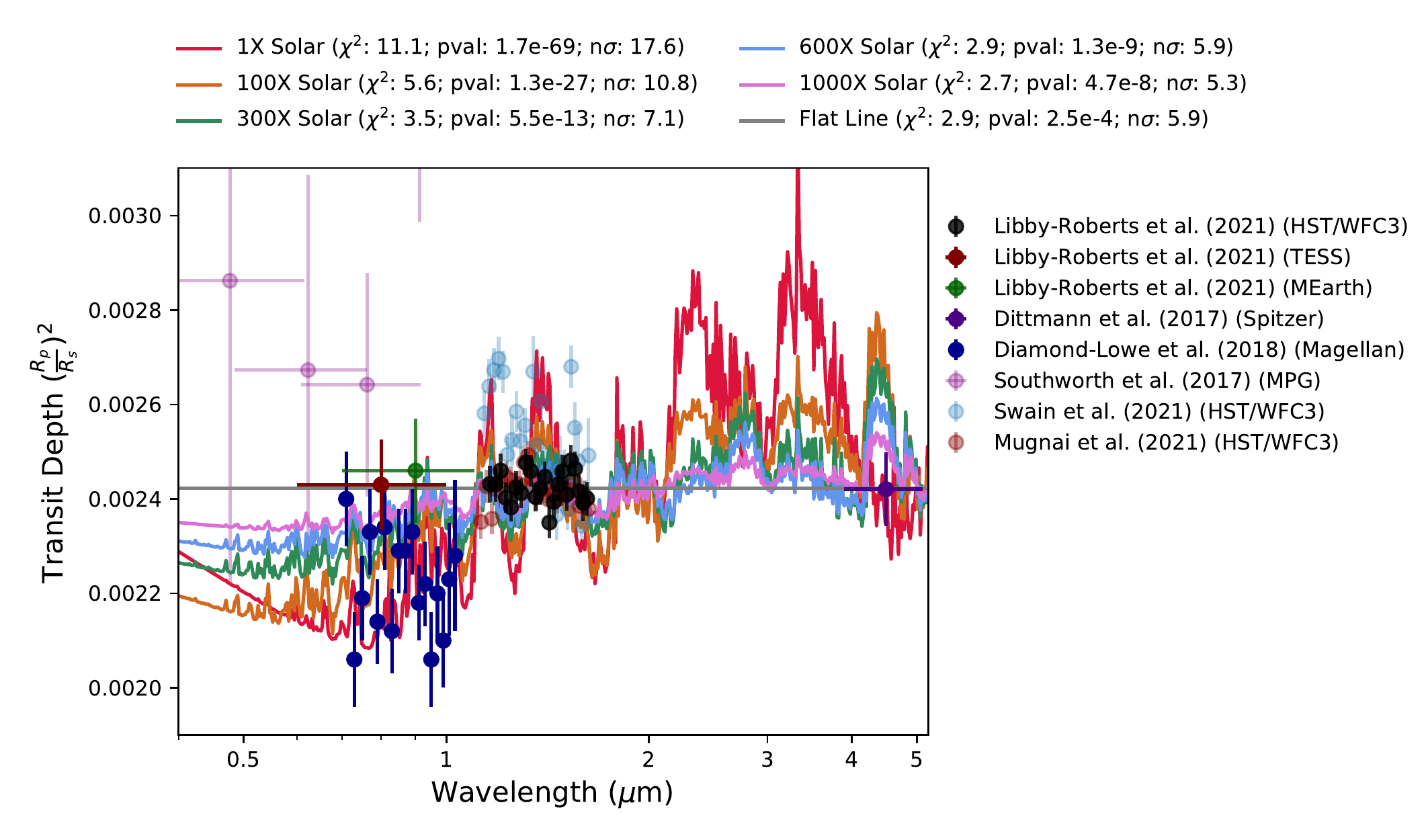}
    \caption{The 0.7--4.5 $\mu$m transmission spectrum of GJ 1132b, combining all available archival data sets with as-published transit depths. The same atmospheric models from Figure~\ref{fig:hst_metal} are included for comparison. Goodness-of-fits for each model were calculated using the LDSS3C spectrum, {\it TESS}, MEarth, and {\it Spitzer} photometry, and the WFC3 depths from this work. The noisy $z-$band flux point \citep{southworth.et.al.2017} was cropped out to ease model visualization on the $y-$axis. From the $\chi^{2}_{r}$ calculated with as-published depths, it appears that no planetary atmosphere model explains all transit depths; however, this is largely driven by the disagreement in the optical LDSS3C spectroscopic depths.}
    \label{fig:spec_no_off}
\end{figure}

\begin{figure}
    \centering
    \includegraphics[width=\textwidth]{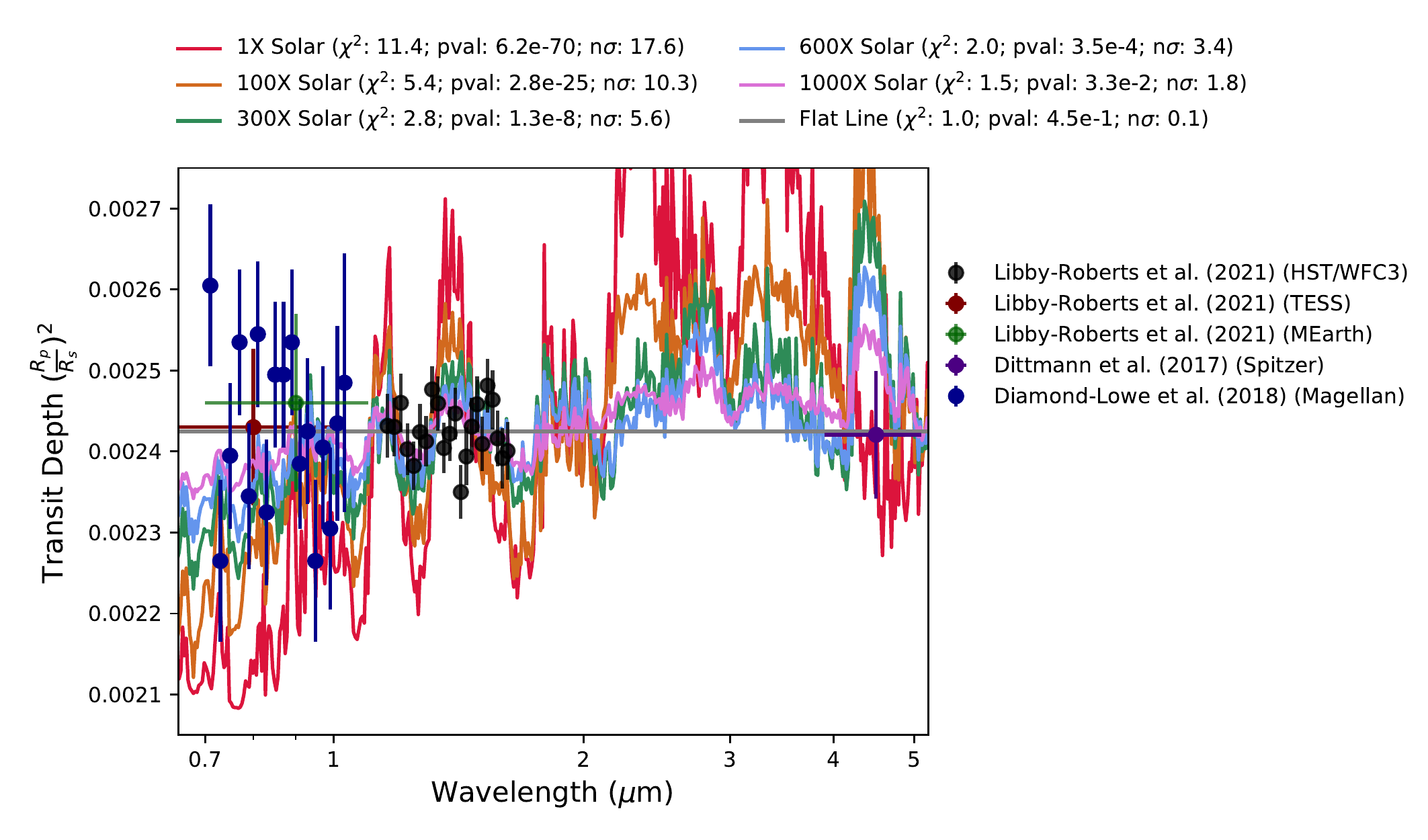}
    \caption{Similar to Figure~\ref{fig:spec_no_off}, but with an offset of 0.000205 added to the \citet{diamondlowe.et.al.2018} spectroscopic transit depths. The addition of archival data strongly supports a featureless transmission spectrum from 0.7--4.5 $\mu$m for GJ 1132b.}
    \label{fig:spec_off}
\end{figure}

A direct comparison between all archival data to the various models shows that every atmospheric model (featureless or otherwise) are ruled out with $>$ 5$\sigma$ confidence (Figure~\ref{fig:spec_no_off}). This is largely driven by the offset of \citet{diamondlowe.et.al.2018} spectroscopic points from the other archival transit depths. Given the discrepancy between the spectroscopic and broadband LDSS3C transit depths, we treat the relative changes in the transit depths within the spectrum as accurate, but not as the absolute depth. We vertically shift these points such that they best fit the different atmospheric models along with the {\it TESS}, MEarth, and {\it Spitzer} broadband points and the WFC3 spectrum (Figure~\ref{fig:spec_off}). We recalculate the goodness-of-fits for all atmospheric models after fitting for this offset. The addition of literature data across 0.7 $-$ 4.5 $\mu$m enables us to rule out 1x, 100x, and 300x Solar metallicities by volume with confidences of 17.6$\sigma$, 10.3$\sigma$, and 5.6$\sigma$ respectively. A flat line remains the best-fit model with a $\chi^{2}_{r}$ of 1.01 ($p$-value: 0.45), while 600x and 1000x Solar metallicity models have $\chi^{2}_{r}$ of 2.0 ($p$-value: 0.00035) and 1.5 ($p$-value: 0.033) respectively. Notably, the {\it Spitzer} 4.5 $\mu$m point disagrees with the 600x Solar metallicity model with $>$2$\sigma$. However, it remains the only observation at wavelengths longer than the 1.63 $\mu$m cutoff from the WFC3 spectrum. Future {\it JWST} observations will be able to fill in this wavelength gap with a precise spectrum.

\section{Starspot Contamination} \label{sec:starspot}
Spatial inhomogeneities on the surface of a star can influence planetary transit depths in a wavelength-dependent fashion, potentially mimicking the signal of absorption through a planetary atmosphere. A planet transiting across a dark spot or a bright facula can show a wavelength-dependent bump \citep{pont.2008.Detectionatmospherichaze, espinoza.2019.ACCESSfeaturelessoptical} that could modify transit depths if not identified and corrected. More perniciously, even a planet transiting an unspotted patch of photosphere will be blocking starlight that is unrepresentative of the average for the star as a whole. This introduces features in observed transit depths (and transmission spectrum) that express the difference between the spotted and unspotted surface \citep{czesla.2009.Howstellaractivity, sing.2011.HubbleSpaceTelescope}. Termed the ``transit light source effect'' (TLSE) by \citet{rackham.2018.TransitLightSource}, this unocculted starspot phenomenon poses a particular challenge for infrared observations of planets transiting cool stellar hosts. M dwarfs have strong molecular features in their photospheres that vary in-and-out of spots, meaning they can introduce spurious transit depth features from molecules like water that might also be expected in a planet's transmission spectrum \citep[see also][]{zhang.2018.NearinfraredTransmissionSpectra, wakeford.2019.DisentanglingPlanetStar}. Here we apply a simple model of the TLSE to the flat WFC3 transmission spectrum in order to infer the starspot properties of the host star GJ 1132 and quantify the impact of the TLSE on current and future observations of GJ 1132b or other similar planets. We outline the modeling framework, enumerate the inputs and assumptions we adopt, and present TLSE constraints.

\subsection{Starspot Model}

\begin{figure}
    \centering
    \includegraphics[width=\textwidth]{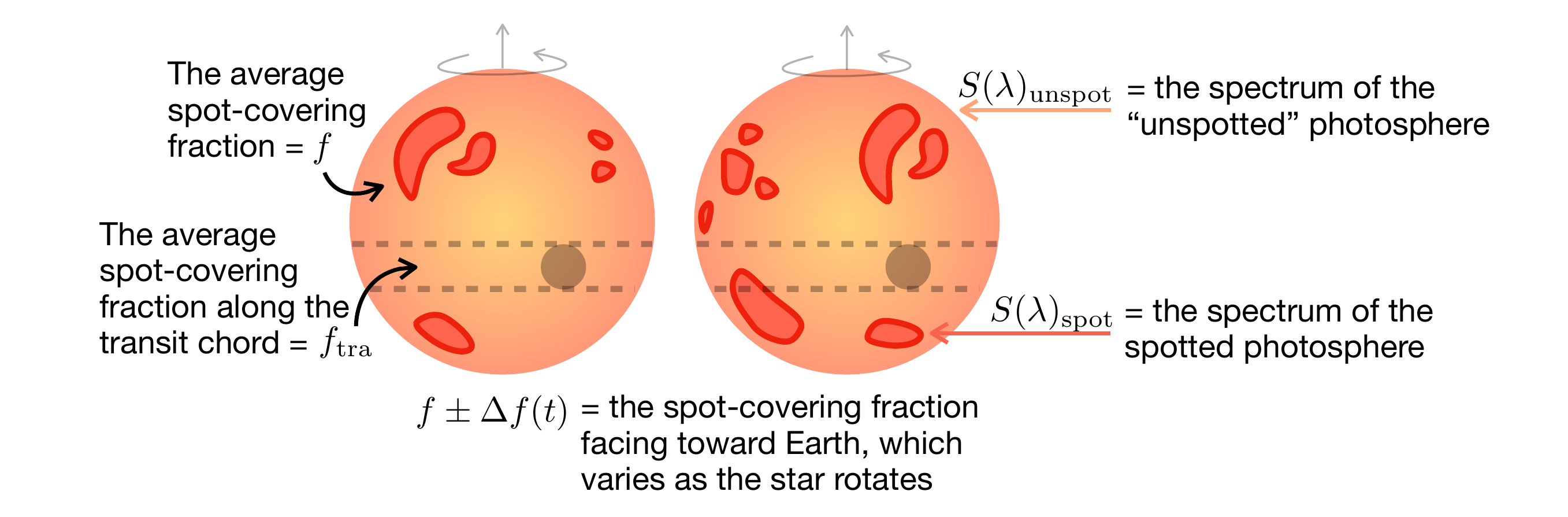}
    \caption{Definition of the variables used for modeling the impact of stellar surface inhomogeneities on the inferred transmission spectrum. In this framework, `spots' can represent patches that are either cooler or hotter than the surrounding stellar photosphere.}
    \label{fig:starspot-variables}
\end{figure}

The wavelength-dependent transit depth of a planet transiting a spotless star can be written as $D(\lambda) = (R_p/R_s)^2 + \Delta D(\lambda)_{\rm atm}$, where the first term represents the transit depth of the planet at a reference radius (such as a rocky surface or a 1-bar level in its atmosphere) and $\Delta D(\lambda)_{\rm atm}=2HR_p/R_s^2 \times n(\lambda)$ represents the absorption by the planet's atmosphere making the planet appear larger by $n(\lambda)$ scale heights at different wavelengths. In this section, we instead aim to model $D(\lambda)$ as an atmosphere-less planet transiting a spotted star. We rewrite the wavelength-dependent depth as $D(\lambda) = (R_p/R_s)^2 + \Delta D(\lambda)_{\rm spot}$, where the spot-induced depth variations ($\Delta D(\lambda)_{\rm spot}$) can be either positive or negative. Approximately following the derivations outlined in \citet{rackham.2018.TransitLightSource} and \cite{zhang.2018.NearinfraredTransmissionSpectra}, the spot-induced depth variation can be estimated as 

\begin{equation}
\Delta D(\lambda)_{\rm spot} =  \left(\frac{R_p}{R_s}\right)^2\times\left[ \frac{(1 - f_{\rm tra}) + f_{\rm tra} \frac{S(\lambda)_{\rm spot}}{S(\lambda)_{\rm unspot}}    }
								 {(1 - f) + f \frac{S(\lambda)_{\rm spot}}{S(\lambda)_{\rm unspot}} } -1\right]
\label{eq:tlse}
\end{equation}
with effectively three unknowns: the flux ratio $S(\lambda)_{\rm spot}/S(\lambda)_{\rm unspot}$, the global spot covering fraction $f$, and the transit-chord spot covering fraction $f_{\rm tra}$, which are  summarized in Figure \ref{fig:starspot-variables} and below.

\begin{itemize}
\item The spectral fluxes $S(\lambda)_{\rm spot}$ and $S(\lambda)_{\rm unspot}$ of the spotted and unspotted photospheres, respectively, represent the flux (in W/m$^2$/nm or related units) emitted by two different types of stellar photosphere. For this work, we use solar-metallicity model stellar spectra from \citet{husser.2013.newextensivelibrary}\footnote{Available for download at \url{https://phoenix.astro.physik.uni-goettingen.de/}} for a surface gravity of $\log g = 5.0$, which closely matches that of GJ 1132, so the temperature ($T_{\rm spot}$ or $T_{\rm unspot}$) uniquely determines the spectrum. To maintain generality for modeling spots or faculae, $T_{\rm spot}$ is allowed to be either cooler or warmer than  $T_{\rm unspot}$. By using fluxes instead of angle-dependent intensities, we are ignoring the effects of limb-darkening but still sensitive to the effect of starspots averaged over the entire stellar disk.

\item The average spot covering fraction $f$ represents the long-term average of the fraction of the Earth-facing stellar disk covered with spots. The instantaneous spot covering fraction facing Earth at any particular time $t$ can be written as $f \pm \Delta f(t)$, where $\Delta f(t)$ represents the variability due to rotation (or, potentially, spot evolution). For example, a star with an average spot covering fraction of $f=20\%$ might exhibit an Earth-facing spot covering fraction ranging from 19\% to 21\% as it rotates through one or more periods. $\Delta f(t)$ therefore has a semi-amplitude of $1\%$. It is important to separate the time-variable term from the average spot covering fraction, as the TLSE depends on the total spot covering fraction in and out of the transit chord, not just the (always smaller) fraction producing out-of-transit modulations. The model for $\Delta D(\lambda)_{\rm spot}$ described here represents the time-averaged impact of starspots on the transit depth rather than the variability from epoch to epoch. This effectively assumes that transits will be observed over multiple epochs and/or that the static signal will dominate over the time-variable contribution to the transit depth. 

\item The transit chord spot covering fraction $f_{\rm tra}$ represents a similar time average along the narrow slice of the stellar disk transited by the planet; the amplitude of $\Delta D(\lambda)_{\rm spot}$ depends on the difference in the spot covering fraction between the disk and the surface actually blocked by the planet. For this work, we assume that $f_{\rm tra} = 0$. Motivated by the lack of obvious spot-crossing features observed in transits, this assumption considerably simplifies Equation \ref{eq:tlse}. However, the $f$ we infer through modeling will more accurately represent how spotted the average disk surface is {\em relative} to the transit chord. It is effectively assuming that the stellar flux along the transit chord can be well represented by a single-temperature PHOENIX atmosphere; the rest of the stellar disk comprises a $f$-weighted average of spectra at two different temperatures.


\end{itemize}

\subsection{Starspot Data and Priors}
To connect this conceptual framework directly to the specific case of GJ 1132b transiting its mid-to-late M dwarf, we apply three data constraints and three priors.

\begin{figure}
    \centering
    \includegraphics[width=\textwidth]{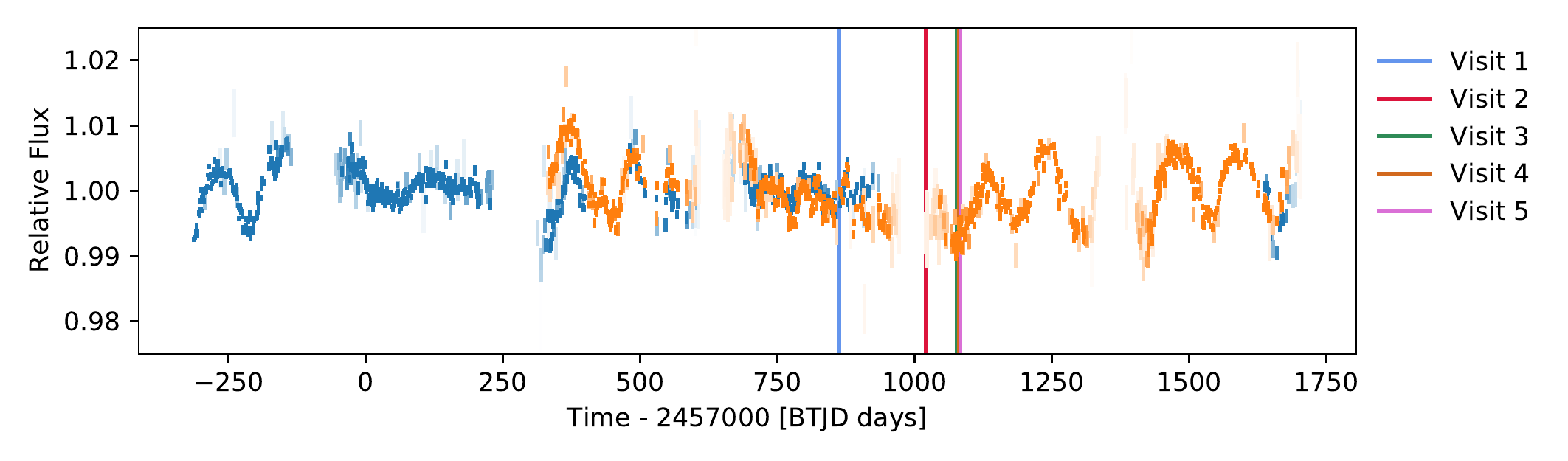}
    \caption{MEarth out-of-transit light curve of GJ 1132, in the MEarth photometric bandpass (roughly 715--1000\,nm), corrected for time-variable precipitable water vapor in Earth's atmosphere. Nightly bins are shown with a color intensity inversely proportional to the nightly variance, for two MEarth telescopes ({\tt tel13} in blue, {\tt tel16} in orange). The epochs of the five WFC3 visits are marked with vertical lines. }
    \label{fig:mearth}
\end{figure}

\paragraph{Data Constraint (1) = The semi-amplitude of out-of-transit modulations} In this model, as a star rotates, the time-variable integrated stellar flux $S(\lambda) + \Delta S(\lambda, t)$ should change with time according to

\begin{equation}
\frac{\Delta S(\lambda, t)}{S(\lambda)} =  -\Delta f(t) \left[ \frac{1 - \frac{S(\lambda)_{\rm spot}}{S(\lambda)_{\rm unspot}}}
{1 - f \left(1-\frac{S(\lambda)_{\rm spot}}{S(\lambda)_{\rm unspot}}\right)}\right],
\end{equation}
which when integrated over any particular bandpass can be compared directly to the semi-amplitude of observed photometric modulations. The star GJ 1132 has been monitored nearly continuously by the MEarth Observatory  \citep{newton.gj1132.rotation} since before the discovery of GJ 1132b. These data trace the photometric variability $\Delta S(\lambda, t)/S(\lambda)$ of the star due to rotation or starspot evolution, integrated over the MEarth RG715 filter bandpass (roughly 700$-$1000\,nm). MEarth out-of-transit monitoring data for GJ 1132, spanning February 2014 through August 2019, are shown in Figure \ref{fig:mearth}\footnote{Available through MEarth Data Release 9 (DR9); \url{https://lweb.cfa.harvard.edu/MEarth/DataDR9.html}}. For the data shown here, all high-cadence transit follow-up observations were excluded. A correction for time-variable precipitable water absorption in Earth's atmosphere was applied using a linear decorrelation against the MEarth common mode \citep[see][]{berta.2012.TransitDetectionMEarth, newton.2016.RotationGalacticKinematics}. Some residual trends are not perfectly corrected through this process, as seen by the disagreements where data from two MEarth telescopes overlap. From the subset of these data that were available at the time, \citet{bertathompson.et.al.2015} estimated a rotation period for GJ1132 of approximately 125 days, \citet{cloutier.2017.RadialVelocityDetection} estimated $122^{+6}_{-5}$ days, and \citet{newton.gj1132.rotation} estimated $130\pm13$ days. The more recent data (Figure \ref{fig:mearth}) continue to support these inferences of a stellar rotation period of roughly 120--130 days. The variation in the light curve's appearance from season to season over 5 years of observations clearly indicates significant starspot evolution and/or differential rotation. We estimate by eye the semi-amplitude of $\Delta S(\lambda, t)/S(\lambda)$ to be about $1.0\pm 0.2$\% in the MEarth bandpass. Though {\it TESS} and {\it Spitzer} also provide precise out-of-transit monitoring, the datasets span durations much shorter than the star's rotation period and therefore cannot be used to estimate the full range of rotational variability of the star. The out-of-transit spectra from this WFC3 program could be used to spectroscopically constrain  $\Delta S(\lambda, t)/S(\lambda)$ at 1.1--1.7\,$\mu$m, but Figure \ref{fig:mearth} shows the five WFC3 epochs do not sample the star at its full range of variability, so we exclude them from the analysis.

\paragraph{Data Constraint (2) = The integrated average stellar flux} In this model, the average integrated spectrum of the star is $S(\lambda) = f S(\lambda)_{\rm spot} + (1 - f)  S(\lambda)_{\rm unspot}$, which is the weighted sum of the spotted and unspotted components. While this model prediction could be compared wavelength-by-wavelength to an absolutely calibrated stellar spectrum, we simplify it into a constraint that the bolometric flux from the star must match that implied by the effective temperature of the star, as
\begin{equation}
\sigma T_{\rm eff}^4 = \int_\lambda S(\lambda) = f \sigma T_{\rm spot}^4 + (1 - f) \sigma T_{\rm unspot}^4
\end{equation}
with $\sigma$ as the Stefan-Boltzmann constant. Without this, arbitrary combinations of spot covering fractions and temperatures could produce stars with extremely unrealistic integrated surface fluxes. For this work we adopt $T_{\rm eff} = 3270\pm140$K \citep{bertathompson.et.al.2015}, implying $\sigma T_{\rm eff}^4 = (6.5\pm1.1)\times10^6$ W/m$^2$ for GJ 1132. 

\paragraph{Data Constraint (3) = The transit depths} In this model, $D(\lambda)$ can be calculated with Equation \ref{eq:tlse} and compared directly to observed transit depth at any wavelengths. For this work, the main transit depths we include are the measured WFC3/G141 transit depths reported in Table \ref{tab:wavedepth}. We also include the spectroscopic LDSS3C depth measurements from \citet{diamondlowe.et.al.2018} because they are the most constraining optical measurements available. As in Section \ref{sec:archival}, we treat the absolute depth of the LDSS3C as unknown. When comparing to starspot models, we include a multiplicative offset allowing these depths to shift up and down together as a free parameter. We marginalize over this offset in all our starspot inferences. These data are shown in Figure \ref{fig:starspot-constraints}, with the maximum-likelihood offset applied to the LDSS3C depths.

\paragraph{Prior (1) = The connection between $\Delta f(t)$ and $f$} In this model, the total spot covering fraction $f$ and the variability in the Earth-facing fraction due to rotation $\Delta f(t)$ are entirely decoupled. However, through a suite of geometric simulations, \citet{rackham.2018.TransitLightSource, rackham.2019.TransitLightSource} determined the amplitude of photometric rotational modulations generally scales as $f^{0.5}$. Such a scaling emerges naturally by treating $\Delta f(t)$ as governed by a Poisson process set by the average number of spots covering the stellar disk. If we define $f_{1}$ as the average spot covering fraction of one individual spot, the number of spots over the entire stellar disk would be $N = f/f_{1}$, and the expected random variation in the number of spots visible at any one time would be $\sqrt{N}$. This implies the expectation value for the level of rotation variability $\Delta f(t)$ should be 
\begin{equation}
\left\langle | \Delta f(t) | \right\rangle = f_{1} \sqrt{N} = \sqrt{f_{1} f}.
\end{equation}
This broadly encapsulates the behavior that high-amplitude rotation variability can emerge from either larger spots (higher $f_{1}$ at fixed $f$) or a more thoroughly spotted star (higher $f$ at fixed $f_{1}$). To capture this relationship, we apply a prior that $x = [f - \Delta f(t)]/f_{1} = N - \Delta f(t)/f_{1}$ should be drawn from a Poisson probability distribution with $N$ as the expectation value, written as 
 \begin{equation}
P(x) = N^{x} e^{-N}/\Gamma(x + 1).
\end{equation} 
By using the gamma function $\Gamma(x+1)$ instead of $x!$ in the denominator, we allow for non-integer values of the number of visible spots. Because we keep $f_{1}$ as a free parameter, this Poisson prior does not actually place new constraints on $f$ and $\Delta f(t)$. Rather, it does allow us to connect them to the typical spot size $f_{1}$ that would best explain the amplitude of rotational variability coming from random Poisson draws for the number of starspots on the Earth-facing side of the star. The inferred typical spot size hinges on the assumption of how starspots are arrayed on the surface.  \citet{rackham.2018.TransitLightSource} assume quasi-isotropic symmetries to estimate rotational modulations scale as $f^{0.5}$. The typical spot size distribution cannot be determined robustly if spots instead congregate to fill active latitudes or polar caps \citep{2018ApJ...868..143G}. 

\paragraph{Prior (2) = Uniform logarithmic priors on $f$, $\Delta f(t)$, $f_1$}
As the values of the spot covering fraction parameters could plausibly span orders of magnitudes, we wish for their priors to be uniform for the logarithm of the value. Therefore, we apply a logarithmic prior of the form $P(x) \propto 1/x$ on each of $f$, $\Delta f(t)$, and $f_{1}$. With this prior, the probability will be uniform across logarithmic intervals, meaning the prior probability of a parameter falling in the range $10^{-3}$--$10^{-2}$ is the same as for the range $10^{-2}$--$10^{-1}$. Furthermore, as the effect of covering $f>0.5$ of the star with cool spots can be reproduced with $f<0.5$ coverage by hot spots, we place a strict prior of $f<0.5$ to avoid those (effectively duplicate) solutions. We also apply a prior requiring $f_{1} < \Delta f(t) < f$.

\paragraph{Prior {3} = The range of allowable spot temperatures} In this model, spot temperatures could be arbitrarily higher or lower than the unspotted surface. However, to avoid introducing an artificial bias toward hot spots due to the practical limitation that the PHOENIX spectra we use are not available below 2300\,K, we impose a symmetrical prior that $0.75 < T_{\rm spot}/T_{\rm eff} < 1.25$. This range  does remove some otherwise viable models, but it still includes most of the spot temperature contrast ratios inferred for cool dwarfs in the \citet{berdyugina.2005.StarspotsKeyStellar}, \citet{2016MNRAS.463.2494F}, and  \citet{morris.2019.StellarPropertiesActive} starspot samples.

\subsection{Starspot Inferences}

We sample from the posterior distribution for the parameters $T_{\rm unspot}$, $T_{\rm spot}$, $f$, $\Delta f(t)$, $f_{1}$, and $R_{p}/R_{s}$ (as well as a nuisance offset for the LDSS3C depths). We use the {\tt emcee} \citep{mcmc.paper} ensemble sampler with 100 walkers. We sample for a total of 20,000 steps and exclude the first 10,000 steps as burn-in. Visual inspection showed that parameter distributions already appear converged by 1000 steps.

\begin{figure}
    \centering
    \includegraphics[width=\textwidth]{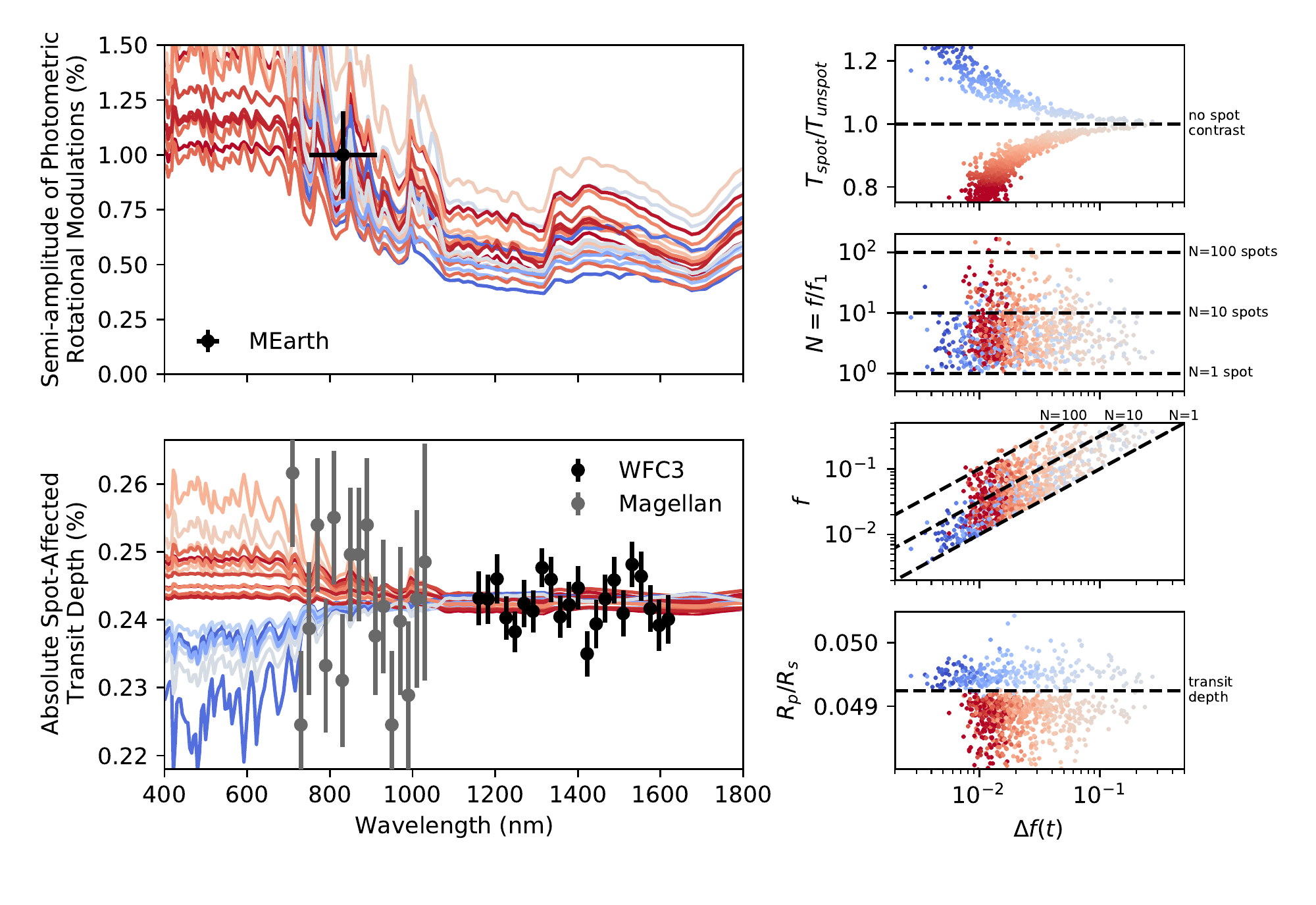}
    \caption{A visualization of the posterior probability distribution for GJ 1132's starspot properties, including data used in the fit ({\em left panels}, with 20 random sample curves) and parameter samples ({\em right panels}, with 1000 random samples). Curves and points are colored from red to blue by $T_{\rm spot}/T_{\rm unspot}$, with paler colors closer to 1. Dashed lines highlight useful reference values in each parameter distribution. A fitted offset was applied to the LDSS3C depths, as explained in the text. The code used to generate this plot is available online (\href{https://github.com/zkbt/gj1132b-starspots-wfc3/blob/main/gj1132b-starspot-constraints-with-wfc3.ipynb}{\texttt{</>}}). }
    \label{fig:starspot-constraints}
\end{figure}

For a first experiment, we assume GJ 1132b's atmosphere has such a high mean molecular weight that its transmission spectrum can be entirely neglected. Thus the planet's radius is effectively constant across wavelength and all transit depth variations are due to the effect of unocculted starspots. Figure \ref{fig:starspot-constraints} visualizes the resulting constraints on GJ 1132's starspot properties. We plot the combinations of fitted parameters that most clearly explain phenomena in the model, and color each sample by the value of its starspot temperature ratio $T_{\rm spot}/T_{\rm unspot}$ so that hotter and colder spots can be traced throughout. We summarize some key conclusions from Figure \ref{fig:starspot-constraints} and this analysis as follows.
\begin{itemize}
    \item The data for GJ 1132 can be closely matched by the starspot model outlined above. At the peak of the posterior distribution, the data yield $\chi^2=46.4$ for 41 data points (1 MEarth semi-amplitude, 1 effective temperature, and 17 LDSS3C and 22 WFC3 depths) and 7 free parameters.
    \item The existing data are insufficient to distinguish whether GJ 1132's most important unocculted surface features (either spots of faculae/plage) are cooler or warmer than the surrounding photosphere. More precise optical data, more reliable knowledge of the absolute depths offset between LDSS3C and WFC3, and/or precise spectroscopic measurements of the rotational modulation semi-amplitude could distinguish between cold or hot spots.
    \item The distribution of $\Delta f(t)$ and $T_{\rm spot}/T_{\rm unspot}$ is shaped by the need to match the semi-amplitude of the rotational modulations; MEarth's observed 1\% variability can be matched either with stronger spot contrasts or more dramatic variations in the spot covering fraction. Only hot spots can extend to $\Delta f(t)<1\%$ and still match the 1\% photometric variability because only they can produce spot flux contrasts $>100\%$, whereas the most that a completely dark cold spot could change a region's surface flux by would be 100\%. 
    \item The distribution of $\Delta f(t)$  and $f$ is shaped additionally by the flat transit depths. $f$ would otherwise uniformly fill the upper left triangle above $f=\Delta f(t)$, but the lack of observed spectral features in the transit depths rules out models with low variability $\Delta f(t)$ (which also have strong temperature contrast) and high average covering fraction $f$.
    \item The average number of visible spots $N = f/f_{1}$ is mostly $<10$, as shown by dashed lines plotted on the distributions for $f$ and $N$. Models with large numbers of very small spots would require too high of $f$ for a given $\Delta f(t)$ and thus introduce strong spectral features into the transit depths. The values of $f = {0.044}_{-0.027}^{+0.094}$ are generally $2-3\times$ the values of $\Delta f(t) = {0.018}_{-0.008}^{+0.029}$. The typical single-spot covering fraction $f_{1} = {0.010}_{-0.006}^{+0.016}$; as a fraction of the $2\pi$ steradians of the Earth-facing side of the star, this would correspond to typical spot radii of $1.1_{-0.8}^{+ 1.3}$ degrees. These sizes are about 2$\times$ larger than the Sun's largest spots at solar maximum \citep{mandal.2017.Kodaikanaldigitizedwhitelight} and similar to the largest active regions recorded on the Sun \citep{hoge.1947.GreatSunspotGroup, nicholson.1948.SunspotActivity1947}. If the planet transited one of these spots at disk center, the crossing event would last about a minute and likely fall entirely within a single WFC3 exposure. We did observe a few single-point outliers in the broadband transit light curves (Figure \ref{fig:four_panel}), but as there were also outliers in the out-of-transit baseline we cannot uniquely attribute them to starspot crossing events.

    \item The true planet-to-star radius ratio $R_{p}/R_{s}$ changes depending on whether the unocculted spots are cold or hot. If the spots are cold, the transit chord is brighter than the rest of the star, so the transit depth over-estimates $R_{p}/R_{s}$ (and vice versa). Marginalizing over both cold and hot spots, we find the true $R_{p}/R_{s} = {0.0490}_{-0.0003}^{+0.0005}$. For GJ 1132b, the flat transit spectrum constrains the spot-induced uncertainties on $R_{p}/R_{s}$ to be broadly similar to the statistical uncertainties on this quantity, and negligible compared to the uncertainties on the stellar radius.
\end{itemize}

The above conclusions connecting rotational variability to the average spot covering fraction rely on an assumption that the longitudinal distribution of spots is random and isotropic. This could be broken by, for example, a large polar spot or an extremely dense zonal band. Such extreme symmetries would exhibit little photometric variability but imbue conspicuous spectral features into the WFC3 depths. We see the opposite (strong photometric variability and flat WFC3 depths), suggesting the Poisson-based approach used here may be valid.

For an additional experiment, we ask whether an unfortunate arrangement of starspots could artificially erase the transmission spectrum of a H/He-rich atmosphere by producing a starspot signal that exactly cancels the extra transit depth from the planet's atmosphere. To answer this question, we produce a fake dataset into which we inject a flipped version of the model transit depths from the above $100\times$ Solar metallicity atmosphere. If the starspot model could explain those flipped depths, the flat observed spectrum could just be a coincidence. We repeat the sampling procedure as above but for simplicity exclude the LDSS3C depths. In this fit, hot spots with strong temperature contrasts and large covering fractions can produce inverted water features that qualitatively mimic those needed to mask the planet's transmission spectrum \citep[as also seen in TRAPPIST-1,][]{zhang.2018.NearinfraredTransmissionSpectra}. However, at the peak of the posterior, $\chi^2 = 59.2$ for 24 data points (1 MEarth semi-amplitude, 1 effective temperature, and 22 WFC3 depths) and 6 free parameters. This is a poor match to the data ($p$-value of $3\times10^{-6}$), mostly driven by the mismatch in shape between the cool water features in the transmission model and the hot water features in the stellar spectra. From this test, we disfavor a scenario where GJ 1132b has a clear H/He-dominated atmosphere that happens to be masked by starspot contamination. 

Another related way of partially cancelling out strong transmission features is the spectral resolution-linked bias \citep{deming.2017.SpectralResolutionlinkedBias}. Transit depths observed at low spectral resolution will be biased toward wavelengths where the star is brightest within each bin. If strong planetary absorption features align in wavelength with deep stellar absorption features, as water lines would do for GJ 1132b and GJ 1132, transmission features will be suppressed. \citet{deming.2017.SpectralResolutionlinkedBias} show the feature suppression to be at the level 12\% for TRAPPIST-1 (500 K cooler than GJ 1132) at WFC3 wavelengths; we do not account for this effect anywhere in our analyses. 

Currently available data suggest that unocculted starspots affect GJ 1132b's transit depths at about the 100 ppm level in the optical and about the 10 ppm level for wavelengths longer than $1\,\mu$m. Starspots do not pose a serious problem for the {\it HST}/WFC3 transmission spectrum of this target, but future analyses of transit observations with {\it JWST} or other large telescopes should be aware of the potential contaminating influence of the star's mottled surface.

\section{Conclusion}\label{sec:conclusion}

We investigated the {\it HST}/WFC3 transmission spectrum of GJ 1132b, a rocky super-Earth orbiting a nearby bright M dwarf, over five separate visits. We determined a featureless spectrum to a precision of 34 ppm in 20 nm wavelength bins spanning 1.15--1.63 $\mu$m. From this result, we rule out the presence of a cloud-free H/He-dominated atmosphere with a mean molecular weight less than 8.9 amu at 4.8$\sigma$ confidence. High-altitude aerosol layers at pressures less than 0.4 mbars could potentially flatten the features of a solar composition atmosphere. Using the predicted mass-loss rate estimated by \citet{waalkes.et.al.2019}, we find that a primordial H/He atmosphere composing 1\% of GJ 1132b's mass could be removed in 3.7 Gyrs by the current high-energy irradiation from the star GJ 1132. With an age $>$5 Gyrs and the much more intense radiation environment it experienced in the past, GJ 1132b is unlikely to possess any primordial H/He dominated atmosphere. We therefore conclude that the most likely scenario for GJ 1132b is that it has a high-metallicity secondary atmosphere, such as Venus' CO$_{2}$-dominated atmosphere, the eroded O$_2$-dominated atmospheres predicted by \citet{schaefer.gj1132b}, or no atmosphere at all. Current data cannot speak to the total mass in such a secondary atmosphere, which could range from a Venusian-like thick atmosphere to Mercurial-like tenuous exosphere. This conclusion is contrary to the recent work by \citet{swain.et.al.2021}, who use a different analysis of the same data set to infer an H/He-rich atmosphere with a Rayleigh scattering slope and HCN and CH$_{4}$ absorption. Our results are consistent with the featureless spectrum found by \citet{mugnai.et.al.2021}. 

We analyzed the GJ 1132 {\it TESS} light curve and determined a broadband transit-depth and orbital parameters for GJ 1132b consistent with those from the WFC3 transits and previous {\it Spitzer} measurements in \citet{dittmann.et.al.2017}. A search for GJ 1132c transits in the {\it TESS} data was unsuccessful, yielding a 3$\sigma$ upper limit of 1.84 R$_{\oplus}$ on the radius of the planet if it transits. Indeed, if GJ 1132c was completely co-planar with GJ 1132b, it would not transit. Instead GJ 1132c requires an orbital inclination of 89.88 degrees (1.2 degrees difference from GJ 1132b) or more for a transit to occur along our line-of-sight. \citet{Bonfils.et.al.2018} note that there is a less than 1\% chance this planet transits, supporting our non-detection in the {\it TESS} light curve.

Combining our GJ 1132b WFC3 spectrum with the {\it TESS} broadband depth, our updated MEarth depth, and other archival GJ 1132b transit depths \citep{dittmann.et.al.2017,diamondlowe.et.al.2018} yielded a transmission spectrum covering 0.7--4.5\,$\mu$m for this planet. We determined that the spectroscopic transit depths from \citet{diamondlowe.et.al.2018} demonstrate a significant offset from the other data sets due to possible instrumental effects. Fitting an offset to the \citet{diamondlowe.et.al.2018} points, we find that the majority of data sets \citep[besides][]{southworth.et.al.2017,swain.et.al.2021} rule out $<$300$\times$ Solar metallicity by volume atmospheric compositions with $>$5.6$\sigma$ confidence. The best-fit model across all points remained a featureless flat line with a $\chi^{2}_{r}$ of 1.01 ($p$-value: 0.45). Future {\it JWST} observations or ground-based observations from the upcoming ELTs will be helpful in discerning the existence of an atmosphere around GJ 1132b and, if one exists, determining its composition.

We explored the influence of unocculted spots on the measured transit depths to assess whether they are able to corrupt the transmission spectrum we infer for the planet. We used a flexible definition for spots as surface features that could be either cool (dark patches of magnetically-suppressed convection) or hot (bright patches like faculae or plage). Given all available data and a Poisson-based model for rotational variability, we estimated that the total spot covering fraction on GJ 1132 is typically 2--3$\times$ larger than the asymmetric distribution of spots that rotates in and out of view and that spot features have typical radii of 0.2--2$^\circ$. We find the likely effect of starspots on GJ 1132b's transit depths to be about 100 ppm in the optical and about 10 ppm in the infrared. As these are comparable to the amplitude of features expected from a secondary atmosphere, future observational studies of GJ 1132b's transmission spectrum should carefully account for unocculted spots.

\acknowledgments

We thank our Program Coordinator Tricia Royle and Contact Scientist Peter McCullough for their efforts to optimize and schedule the WFC3 observations, as well as the entire team that makes {\it Hubble} the telescope that it is. We also thank Will Waalkes, Girish Duvvuri, and Carlos Cruz-Arce for their helpful conversations and suggestions on this project. We gratefully acknowledge that this work is based on observations made with the NASA/ESA Hubble Space Telescope, obtained from the data archive at the Space Telescope Science Institute (STScI) and supported by NASA through grant numbers HST-GO-14758 and HST-AR-15788 from STScI. STScI is operated by the Association of Universities for Research in Astronomy, Inc. under NASA contract NAS 5-26555. This material is also based upon work supported by NASA under grant No. 80NSSC18K0476 issued through the XRP Program. This publication was made possible through the support of a grant from the John Templeton Foundation. The opinions expressed here are those of the authors and do not necessarily reflect the views of the John Templeton Foundation. Z.K.B.-T. is thankful for support from the National Science Foundation NSF/CAREER program under Grant 1945633. B.V.R. thanks the Heising-Simons Foundation for support. J.M.D acknowledges support from the Amsterdam Academic Alliance (AAA) Program, and the European Research Council (ERC) European Union’s Horizon 2020 research and innovation programme (grant agreement no. 679633; Exo-Atmos). This work is part of the research programme VIDI New Frontiers in Exoplanetary Climatology with project number 614.001.601, which is (partly) financed by the Dutch Research Council (NWO).

%

\vspace{5mm}
\facilities{HST(WFC3), TESS, MEarth}


\software{Astropy \citep{astropy.2013, astropy.2018},  
         Photutils \citep{photutils},
         emcee \citep{mcmc.paper},
         BATMAN \citep{batman.paper},
         LDTK \citep{ldtk.2015},
         Iraclis \citep{iraclis.2016a,iraclis.2016b},
         RECTE \citep{zhou.et.al.2017},
         Exo-Transmit \citep{exotransmit.cite},
         Lightkurve \citep{lightkurve.cite}
          }




\bibliography{references}

\end{document}